# Upconverting Core-Shell Nanocrystals with High Quantum Yield under Low Irradiance: On the Role of Isotropic and Thick Shells

Stefan Fischer[1*], Noah J.J. Johnson[2], Jothirmayanantham Pichaandi[2], Jan Christoph Goldschmidt[1], and Frank C.J.M van Veggel[2*]

[1]*Fraunhofer Institute for Solar Energy Systems, Heidenhofstraße 2, 79110 Freiburg, Germany*
[2]*Department of Chemistry, University of Victoria, P.O. Box 3065, Victoria, British Columbia, Canada V8W 3V6*
*sfischer.public@gmail.com; fvv@uvic.ca

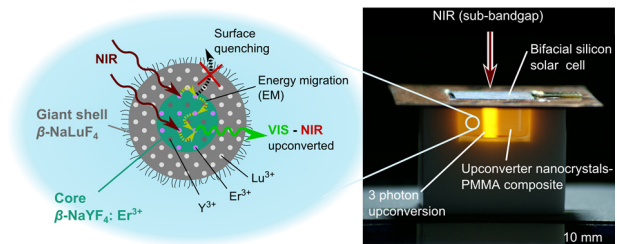

**Abstract** Colloidal upconverter nanocrystals (UCNCs) that convert near-infrared photons to higher energies are promising for applications ranging from life sciences to solar energy harvesting. However, practical applications of UCNCs are hindered by their low upconversion quantum yield (UCQY) and the high irradiances necessary to produce relevant upconversion luminescence. Achieving high UCQY under practically relevant irradiance remains a major challenge. The UCQY is severely limited due to non-radiative surface quenching processes. We present a rate equation model for migration of the excitation energy to show that surface quenching does not only affect the lanthanide ions directly at the surface but also many other lanthanide ions quite far away from the surface. The average migration path length is in the order of several nanometers and depends on the doping as well as the irradiance of the excitation. Using $Er^{3+}$-doped $\beta$-NaYF$_4$ UCNCs, we show that very isotropic and thick (~10 nm) $\beta$-NaLuF$_4$ inert shells dramatically reduce the surface-related quenching processes, resulting in much brighter upconversion luminescence at simultaneously considerably lower irradiances. For these UCNCs embedded in PMMA, we determined an internal UCQY of 2.0±0.2 % using an irradiance of only 0.43±0.03 W/cm$^2$ at 1523 nm. Normalized to the irradiance, this UCQY is 120× higher than the highest values of comparable nanomaterials in the literature. Our findings demonstrate the important role of isotropic and thick shells in achieving high UCQY at low irradiances from UCNCs. Additionally, we measured the additional short-circuit current due to upconversion in silicon solar cell devices as a proof concept and to support our findings determined using optical measurements.

## 1. Introduction

Upconverter materials absorb two or more photons and subsequently emit one photon with higher energy than that of each of the absorbed ones. Colloidal upconverter nanocrystals (UCNCs) are relevant for a variety of different applications including theranostics,[1-3] bioimaging,[4-6] and photovoltaics.[7, 8] Increasing the internal upconversion quantum yield (UCQY) - the ratio of emitted higher energy photons to the absorbed photons - leads to stronger upconversion (UC) luminescence at a given irradiance of the excitation. Therefore, in bioimaging and theranostics, a higher UCQY improves the signal-to-noise relation due to brighter upconverter samples.[2] The UCQY generally increases with the irradiance owing to the non-linear nature of the upconversion processes. However, in bioimaging, high excitation intensities introducing local heating and phototoxic effects need to be avoided to study living cells.

Hence, more efficient upconverter samples are required, which are brighter at lower excitation intensities. In photovoltaics, achieving a high UCQY at appropriately low irradiance, typically in the range of several 0.1 W/cm$^2$, is a mandatory precondition to achieve a significant impact of the upconverter on the overall device performance.[9]

Here, the upconverter material converts the otherwise unused sub-bandgap photons, which carry about 20% of the incident solar power in the case of crystalline silicon solar cells, into photons with energies above the bandgap.[10] These photons can then be utilized by the solar cell (Figure 1a). Theoretically, the efficiency of a crystalline silicon solar cell can be enhanced by UC from 30% to 40% (33% relative).[10] In more comprehensive studies, considering up to data and more realistic parameters for silicon solar cells and $Er^{3+}$-doped upconverter materials, the predicted efficiency enhancement is around 15% relative.[11, 12]

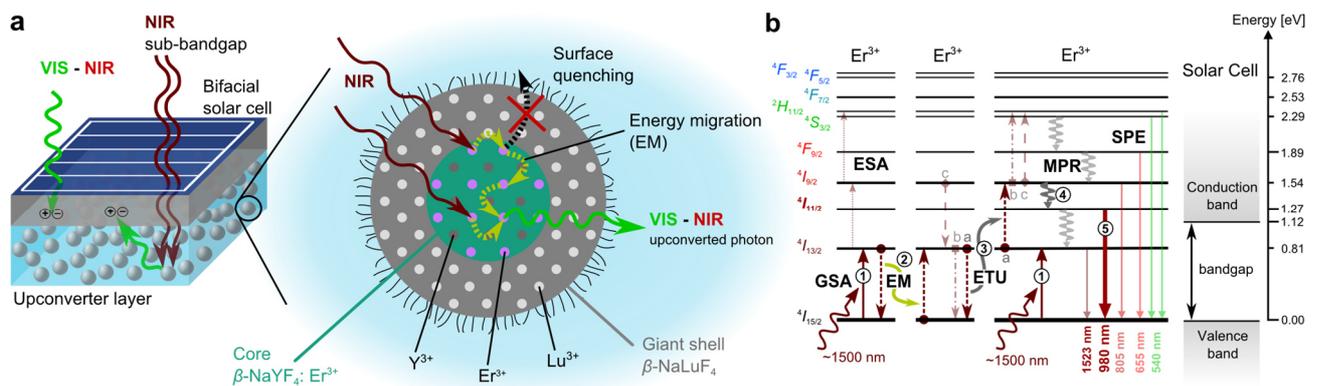

**Figure 1.** (**a**) NIR sub-bandgap photons are transmitted through the solar cell and are absorbed by the upconverter nanocrystals (UCNCs). Two or more photons are converted to one photon that can then be utilized by the solar cell resulting in a higher efficiency of the photovoltaic device. The optically active core of the nanocrystals can be covered with a thick and isotropic passivating shell. This thick and isotropic shell reduces considerably the typically strong surface quenching, which leads to much brighter and more efficient UCNCs. (**b**) Schematic representation of the upconversion processes for an $Er^{3+}$-based upconverter. The most efficient upconversion process is ground state absorption (GSA) followed by energy transfer upconversion (ETU) between neighboring $Er^{3+}$ ions both in an excited state. Additionally, energy migration (EM) due to excitation hopping from one ion to another ion has to be taken into account.

Trivalent erbium ($Er^{3+}$) doped upconverter materials are the most efficient ones with respect to upconversion for crystalline silicon solar cells, where photons with wavelengths longer than roughly 1200 nm need to be converted to shorter wavelengths and consequently higher energies.[13-16] Typically, upconverter materials doped solely with $Er^{3+}$ that are excited by photons with a wavelength around 1500 nm, show a dominant UC emission with a center wavelength around 980 nm, which can then be utilized efficiently by a silicon solar cell[13, 17-19], as illustrated in Figure 1a,b.

Since the UCQY typically increases with the irradiance, one important boundary condition for upconversion in the context of photovoltaics is the solar radiation available for upconversion. While the total irradiance from the sun on the surface of the earth is around 0.1 W/cm$^2$, the sun provides an irradiance of less than 0.0024 W/cm$^2$ in the absorption range of $Er^{3+}$-doped upconverter materials from 1480 nm to 1580 nm, using the global standard solar spectrum with air mass 1.5 (AM1.5g).[20] This value could be increased by external concentration of the solar radiation to a maximum possible irradiance of 110 W/cm$^2$, using the theoretical limit for the solar concentration factor of 46,200×.[21] In concentrator photovoltaics, however, the solar radiation is typically only concentrated by a factor of 300 to 1000× with a corresponding irradiance suitable for $Er^{3+}$ upconverter materials of 0.72 to 2.40 W/cm$^2$.

The reported efficiency enhancement due to upconversion for bifacial crystalline silicon solar cells are far below the theoretical limit. The highest reported values are <1% relative using solar concentration factors around 100×.[16, 22] To realize the full potential of upconversion to enhance the solar cell's efficiency, additional means are clearly necessary such as (i) widening the narrow absorption range of the rare-earth ions by a second luminescent material, which acts as a broadband absorption sensitizer [8, 23] and (ii) incorporation of the upconverter into a photonic structure or exploiting plasmonic effects near metal surfaces for local irradiance enhancement and spontaneous emission control.[24-26] However, these concepts command the use of nanomaterials because the upconverter needs to be integrated into nanostructures and be placed at specific positions in the nanometer range to achieve the desired effects.

The goal of a relevant efficiency enhancement in any application, however, requires that the UCNCs exhibit a high UCQY to start with. Unfortunately, the performance of UCNCs is orders of magnitude lower than that of the bulk counterparts.[27-29] For bulk $\beta$-NaYF$_4$: 25% $Er^{3+}$ (microcrystalline powder) the reported values for the internal UCQY are as high as 12.0±1.0 % using a low irradiance of 0.40±0.02 W/cm$^2$ at 1523 nm.[30] In contrast, the highest reported value for comparable UCNCs can be calculated from the data given by Shao et al. [31] (see Supporting Information). They investigated core-shell UCNCs consisting of $\beta$-NaYF$_4$: 10% $Er^{3+}$ cores with diameters of 22.0±0.7 nm and elliptically shaped inert $\beta$-NaYF$_4$ shells with average short axes of 25.0±0.5 nm and long axes of 38.0±1.0 nm. The internal UCQY was calculated to be 0.7% under 1523 nm monochromatic illumination using an irradiance of 18 W/cm$^2$,[31] which is well above the value that can be achieved with solar radiation using geometrical concentration optics for photovoltaics. Hence, for the UCNCs a more than one order of magnitude *lower* internal UCQY is found at simultaneously several orders of magnitude *higher* irradiance levels than their bulk counterparts. This is also reflected in an often-used figure of merit, where the internal UCQY is divided by the irradiance.[13, 14, 18, 30] This normalized UCQY decreases for increasing irradiances.[14, 16] Although the normalized UCQY is thus not independent of the irradiance, it helps to evaluate the UCQY values of upconverter materials, which were determined at very different irradiances. The normalized UCQY of the above discussed materials is 0.3 cm$^2$/W for the bulk material while for the nanomaterials the highest value is only 3.9×10$^{-4}$ cm$^2$/W. This illustrates the very poor performance of UCNCs compared to the bulk counterparts, which is only 1/750 of the bulk UCQY value.

Several approaches to enhance the performance of UCNCs have been discussed.[29] Co-doping with different lanthanide ions and active shell concepts have been investigated to broaden the absorption range of upconverting nanocrystals.[32] The predominant approach has been the growth of an inert outer shell.[27-29, 33-35] This shell shields the optically active core from surface quenching[27-29], as illustrated in Figure 1a. However, surface-related quenching is still the main limiting factor of the low UCQY in core-shell nanocrystals owing in part to a poor integrity of the shell layers or a too thin passivating shell.[36, 37]

In this article, we demonstrate that UCQY values comparable to those of bulk materials can be achieved for nanocrystals at low irradiance levels applicable to photovoltaics. For this purpose, we synthesized an isotropic, yet very thick epitaxial shell around the optically active core.[38, 39] We performed a comprehensive study on the UCQY as a function of the irradiance on these novel core-shell UCNCs with very thick and isotropic passive shell (CS2) both in chloroform (CHCl$_3$) and embedded in poly(methyl methacrylate) (PMMA). Static surface quenching was reduced dramatically due to the 10 nm thick shell, which led to much higher UCQY values as determined for reference samples with thinner or no shell. Using a simplified rate equation model for energy migration of the excited states, we elucidate the dynamics of the upconversion process and the relevance of a complete surface passivation by an isotropic shell. Finally, we applied the embedded nanocrystals to an optimized bifacial silicon solar cell to build the first upconverter crystalline silicon solar cell device using UCNCs. The external quantum efficiency (EQE) due to upconversion of sub-bandgap photons was measured to demonstrate and confirm the high UCQY of the UCNCs for potential application in photovoltaics.

## 2. Experimental Section

### A. Synthesis and material characterization of the core-shell upconverting nanocrystals

Core-shell nanocrystals have been synthesized using the hot injection method as reported by Johnson *et al.* in Ref. 38. The (cubic) $\alpha$-NaLuF$_4$ sacrificial nanocrystals (SNCs) were synthesized following a reported procedure and dissolved in hexane.[40] The active core $\beta$-NaYF$_4$: Er$^{3+}$ nanocrystals have been synthesized as reported in Ref. 41. Approximately 0.6 mmol of the $\alpha$-NaLuF$_4$ SNCs were added into vials, the hexane evaporated until a volume of approximately 0.5 ml was left and then mixed with 1 ml of 1-octadecene. Such a mixture was taken with a syringe and injected into the hot (300 °C) reaction vessel containing the nanocrystals to grow a shell layer. Injections were repeated after a ripening time of 10 minutes. For the CS2 UCNCs, injections of the $\alpha$-NaLuF$_4$ SNCs were repeated 9 times. For the CS1 UCNCs approximately 1.5 mmol of the SNCs were deposited on the $\beta$-NaYF$_4$: Er$^{3+}$ core with 3 injections. After the ripening time of the last injection the solutions were cooled down to room temperature. The nanocrystals were precipitated by addition of ethanol, collected by centrifugation, washed twice with ethanol and finally dispersed in hexanes. Later the UCNCs were transferred to chloroform (CHCl$_3$) by centrifugation, subsequent evaporation of the residual hexane under vacuum with a rotary evaporator (Hei-VAP Precision, Heidolph), and finally re-dispersion in CHCl$_3$.

The TEM images were captured using a JEOL JEM-1400 microscope operating at 80 kV. The size of the core and core-shell nanocrystals was determined from the TEM images by counting more than 100 nanocrystals.

The mass of the inorganic content of the UCNCs in a certain volume of solvent was determined with TGA (STA 449 F3 Jupiter, Netzsch) at a heating rate of 5 °C per minute from room temperature to 550 °C, after evaporating the solvent.

The concentration of nanocrystals in the solution was estimated with the initial sample mass in the aluminum crucibles, the inorganic content measured with TGA, and the size of the nanocrystals determined from the TEM images assuming spherical particles. This concentration of nanocrystals was also used to calculate the concentration of nanocrystals in the composite samples. The number of Er$^{3+}$ per nanocrystal and the Er$^{3+}$ concentration per volume was estimated from the core size, the nominal Er$^{3+}$ doping, and the concentration of the nanocrystals in the solution.

All chemicals were purchased from Sigma-Aldrich except oleylamine (97%) from Acros, anhydrous ethanol from Commercial Alcohols, methanol from Caledon, and Lu$_2$O$_3$ from MV Laboratories. All chemicals were used as received.

### B. Embedding of Nanocrystals in PMMA

We embedded the UCNCs in PMMA by radical polymerization using the procedure reported in Ref. 42. In a typical fabrication, an appropriate amount of nanocrystals was added into a 10 ml glass vial with an inner diameter of 18 mm. Approximately 80 ml poly(ethylene glycol)-monooleate (PEG) with a molecular weight of $M_n$ = 860 g/mol was added into the vial and the CHCl$_3$ was thoroughly removed under vacuum with a rotary evaporator (Hei-VAP Precision, Heidolph). 3.4 g of methyl methacrylate (MMA) was added to the vial and the dispersion was sonicated in a bath sonicator until a clear solution was obtained. Subsequently, 10 mg of the radical initiator azobisisobutyronitrile (AIBN) was added and the solution sonicated once more for approximately 1 minute. The vials containing the monomer solution were sealed and immersed in a 70 °C silicone oil bath for 30 minutes. Afterwards, the silicone oil bath was cooled down to 50 °C until the polymerization was completed. Translucent nanocrystals-PMMA composite cylinders were obtained.

## C. Method and Determination of the Upconversion Quantum Yield (UCQY)

We used the experimental setup for the spectroscopic measurements as well as the methodology to determine the UCQY values as reported in the literature.[9, 15] We distinguished between two definitions for the UCQY: the *internal UCQY* and the *external UCQY*. The external UCQY (*eUCQY*) is defined as the photon flux of upconverted photons emitted by the upconverter with more energy than the energy of the photons from the excitation $\phi_{UC}$ divided by the photon flux of the photons incident on the upconverter sample $\phi_{in}$ via

$$eUCQY = \frac{\text{upconverted photon flux}}{\text{incident photon flux}} = \frac{\phi_{UC}}{\phi_{in}}. \quad (1)$$

The external UCQY describes how efficient the incident light is upconverted. Therefore, it is the more relevant quantity in most applications. The photon flux density of the emitted upconverted photons (brightness of the upconversion) is determined by the external UCQY multiplied by the incoming photon flux density of the excitation, which is proportional to the irradiance of the excitation. Therefore, the ultimate performance and figure of merit of the UCNCs is well and completely described by the external UCQY.

However, in the literature more often values of the internal UCQY (*iUCQY*) are given, which is the ratio of the photon flux of upconverted photons $\phi_{UC}$ divided by the photon flux of absorbed photons $\phi_{abs}$:

$$iUCQY = \frac{\text{upconverted photon flux}}{\text{absorbed photon flux}} = \frac{\phi_{UC}}{\phi_{abs}}$$
$$= \frac{1}{A_{UC}} eUCQY. \quad (2)$$

The internal UCQY reflects how efficient the physical process of upconversion itself is. In consequence, the internal UCQY is independent of the density of UCNCs in the solutions as long as other parasitic effects, such as reabsorption of emitted upconverted photons, can be neglected. Following this definition, for a two-photon upconversion process the internal UCQY must be ≤50%. The absorptance of the sample $A_{UC}$ connects the internal UCQY with the external UCQY. Hence, increasing the absorptance of the sample, for example by a larger amount of UCNCs in solution or the polymer, enhances the external UCQY. Eventually, the external UCQY will approach the internal UCQY for 100% absorptance.

All samples were mounted inside an integrating sphere (819C-SL-5.3, Newport). All instruments as well as the complete setup were calibrated. Further details and a figure of the setup can be found in the Supporting Information. The integrating sphere features two fiber ports. One of the attached optical fiber bundles, with a round to angular shape, guides the light directly to a spectrometer (SP2300i, Princeton Instruments) equipped with a Si-CCD detector (PIXIS: 256E, Princeton Instruments). The used grating has a blaze wavelength of 800 nm and 150 lines/mm. We measured a spectral resolution of less than 1 nm for this system. The entrance slit had a width of 50 µm. The other optical fiber guides the light to a monochromator (H25, Jobin Yvon) equipped with a gold coated grating with a blaze wavelength of 1000 nm with 600 lines/mm. Attached to the monochromator is a stack detector, consisting of a silicon (Si) photodiode on top of an indium gallium arsenide (InGaAs) one (OEC GmbH). Hence, the monochromator unit is optimized for the NIR and was predominantly used to measure the laser spectrum. The absorptance was determined by the difference of the integrated laser signal with the upconverter samples and the un-doped reference sample inside the integrating sphere. Furthermore, a calibrated Si photodiode (818-SL-L, Newport) was directly applied to the integrating sphere to measure the integrated emission from the upconverter samples up to approximately 1100 nm. The external quantum efficiency of the Si photodiode was determined with an uncertainty of less than 1% absolute. The signal from the Si photodiode was measured via a preamplifier using a voltmeter (HP 34401A, Agilent).

The collection efficiency $\beta$ of the integrating sphere using the Si photodiode was determined with a calibrated tungsten halogen lamp. The relative emission spectrum of the halogen lamp was precisely known from calibrated measurement using another calibrated setup. The photon flux into the integrating sphere was measured for different calibration cycles with different absolute photon fluxes. The photons flux from the tungsten halogen lamp was altered with different pin holes before the light from the tungsten halogen lamp could enter into the integrating sphere. The collection efficiency $\beta$ was determined for more than 10 calibration cycles and a standard deviation of less than 5% was determined. Using the same tungsten halogen lamp the spectral response (correction function) of the spectrometer and the monochromator were determined. All spectra were corrected with the corresponding spectral response of the instruments.

The collimated laser beam from a tunable NIR laser (TSL-510, Santec) was used for the excitation of the samples. A Gaussian laser beam profile was measured close to the sample position using an InGaAs CCD camera (Xeva InGaAs, Xenics). We used the full area at half maximum (FAHM) to define the area of the laser beam $A_{laser}$. For an ideal Gaussian peak this corresponds to the full width at half maximum (FWHM). Very often the $1/e$ value or the $1/e^2$ value of the peak laser power is used to define the diameter of the laser beam. The beam diameter using the $1/e^2$ value is 1.7 times larger than the one for the FWHM value. Consequently, the irradiance on the sample is much lower for the $1/e^2$ definition compared to the FWHM definition. We used the FWHM because it is a more conservative definition compared to others. Because UC is a non-linear process, the definition of the beam area makes a crucial difference when comparing the UCQY values of materials.

For every sample, the upconversion luminescence as well as the laser signal was measured at least 10 times, whereby the samples were measured alternately – meaning one sample was taken out of the integrating sphere after one measurement and another sample was introduced. The error on the UCQY values was calculated by Gaussian error propagation of the random error from the repeating measurements as well as the systematic error from the setup calibration.

### D. Upconverter Solar Cell Device Measurements

The CS2 UCNCs-PMMA composites were applied on the rear side of a bifacial silicon solar cell with an index matching liquid (Cargille, Type 300) and placed onto a PTFE measurement chuck as described in Ref. 15.

We measured the short-circuit current of the solar cell due to upconversion. The same NIR laser was used as for the optical measurements with an excitation wavelength of 1523 nm. The external quantum efficiency (EQE) of the solar cell due to the upconversion of sub-bandgap photons, denoted as $EQE_{UC}$, was determined using a calibrated germanium reference solar cell as reported in Refs 9, 15.

For a direct comparison of the electrical and the optical measurement, the $EQE_{UC}$ has to be corrected for the incomplete transmittance of the excitation through the solar cell $T_{cell}$ and the efficiency by which the upconverted photons are utilized at the rear side of the solar cell $EQE_{rear}$.[16] The $EQE_{rear}$ for a wavelength of 980 nm was used because more than 98% of the upconversion luminescence originates from corresponding transition $^4I_{11/2} \rightarrow {^4I_{15/2}}$. The corrected $EQE_{UC}$, which can directly be compared to the optically determined external UCQY, is denoted as $UCQY_{elec}$ and was calculated by

$$UCQY_{elec}(T_{cell} \, I) = EQE_{UC}(I) \frac{1}{EQE_{rear}(980 \, nm)} \quad (3)$$

## 3. Energy Migration Model

The excitation energy of a donor $Er^{3+}$ ion (energy state $^4I_{13/2}$) can migrate through the crystal by diffusion or hopping. In our case, the hopping is mediated by Förster resonant energy transfer ($^4I_{13/2}, {^4I_{15/2}} \rightarrow {^4I_{15/2}}, {^4I_{13/2}}$) between neighboring $Er^{3+}$ ions, which is highly efficient due to the large overlap of the corresponding emission and absorption spectra from the same transition ($^4I_{13/2} \leftrightarrow {^4I_{15/2}}$).[43-46] The hopping model is typically valid as $C_{DD} \geq C_{DA}$.[47] Here $C_{DD}$ and $C_{DA}$ are parameters describing the strength of energy transfer from one donor to another donor (DD) and from one donor to an acceptor (DA), respectively. In our case, the acceptor is another $Er^{3+}$ ion also in the $^4I_{13/2}$ energy state that is further excited by energy transfer upconversion into the $^4I_{9/2}$ energy state. Hopping of the excitation energy in rare-earth ions has been extensively investigated.[43, 46] The probability for energy transfer due to dipole-dipole interaction for a donor-donor process can be described by

$$P_{ET,DD}(d) = \frac{1}{\tau_D} \left( \frac{R_{0,DD}}{d} \right)^6 = \frac{C_{DD}}{d^6} \quad (4)$$

and accordingly for donor-acceptor process

$$P_{ET,DA}(d) = \frac{1}{\tau_D} \left( \frac{R_{0,DA}}{d} \right)^6 = \frac{C_{DA}}{d^6} \quad (5)$$

using the lifetime of the donor state $\tau_D$, the critical Förster radius $R_{0,DD}$ for the donor-donor energy transfer ($^4I_{13/2}, {^4I_{15/2}}) \rightarrow ({^4I_{15/2}}, {^4I_{13/2}})$, the critical Förster radius $R_{0,DA}$ for the donor-acceptor energy transfer ($^4I_{13/2}, {^4I_{13/2}}) \rightarrow ({^4I_{15/2}}, {^4I_{9/2}})$, respectively, and the average Er-Er distance $d$. Some literature values of $\tau_D$, $C_{DD}$, and $C_{DA}$ are given in Table 1.

**Table 1.** Literature values of $\tau_D$, $C_{DD}$ ($^4I_{13/2}, {^4I_{15/2}}) \rightarrow ({^4I_{15/2}}, {^4I_{13/2}})$, and $C_{DA}$ ($^4I_{13/2}, {^4I_{13/2}}) \rightarrow ({^4I_{15/2}}, {^4I_{9/2}})$.

| $\tau_D$ [ms] | $C_{DD}$ [cm$^6$/s] | $C_{DA}$ [cm$^6$/s] | Material System | Ref. |
|---|---|---|---|---|
| 10.00 | 4.1×10$^{-39}$ | - | LiYF$_4$:Er$^{3+}$ | 44 |
| 8.85 | 2.4×10$^{-39}$ | 4.7×10$^{-39}$ | LiYF$_4$:Er$^{3+}$ | 45 |
| 7.55 | 5.0×10$^{-39}$ | 1.6×10$^{-39}$ | Al$_2$O$_3$:Er$^{3+}$ | 46 |

In the following, we describe how we determine the average migration path starting from the donor ion that is first excited due to absorption of a photon. Therefore, we calculate (i) the number of average hopping processes from one $Er^{3+}$ to another $Er^{3+}$ and (ii) from the number of hopping processes and the average Er-Er distance we determine the average migration distance based on a random walk model.

The number of hopping processes $n_{hop}$ can be estimated in a very simple way by the ratio $\tau_D/\tau_0$, where $\tau_D$ is the lifetime of the donor state and $\tau_0$ is the average time an $Er^{3+}$ excitation resides before it is hopping to another $Er^{3+}$. The average time $\tau_0$ can be determined by

$$\tau_0 = \left(\frac{3}{2\pi}\right)^3 \frac{1}{N_D^2 C_{DD}} . \quad (6)$$

as reported in Refs. 47, 48. In our case, the density of donors $N_D$ equals the density of $Er^{3+}$. Using the corresponding literature values for the donor lifetime the $n_{hop}$ values are in the range of a few thousands (see also Supporting Information).

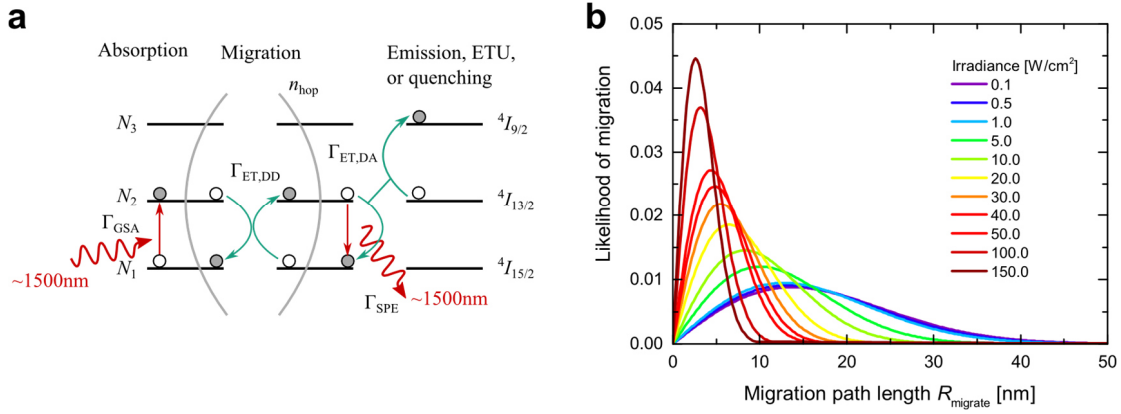

**Figure 2.** (**a**) Scheme of the rate equation model for migration of the donor excitation ($^4I_{13/2}$) by energy transfer (hopping, $^4I_{13/2}, ^4I_{15/2} \rightarrow {}^4I_{15/2}, ^4I_{13/2}$). After a number of hopping processes ($n_{hop}$) the excited ion relaxes back to the ground state $^4I_{15/2}$ or finds another ion in an excited state for ETU. (**b**) The likelihood of the migration path length $R_{migrate}$ was calculated by combining the migration rate equation model with a random walk model. Shorter $R_{migrate}$ are found for higher irradiances due to the higher population density of excited ions, which results in a higher probability of ETU.

Using a simplified rate equation model is another and more sophisticated way of calculating $n_{hop}$. In general, the probability for one hopping step to occur is determined by the probability of energy transfer (hopping, $^4I_{13/2}, ^4I_{15/2} \rightarrow {}^4I_{15/2}, ^4I_{13/2}$) in relation to the probability of all other process, such as spontaneous emission, energy transfer upconversion ($^4I_{13/2}, ^4I_{13/2} \rightarrow {}^4I_{15/2}, ^4I_{9/2}$), multi-phonon relaxation, and possible non-radiative recombination due to defects in the volume. A scheme of the simplified model we used is depicted in Figure 2a. The model is based on a 3 energy level system. The probability for one hopping process after an $Er^{3+}$ ion is first excited into the $^4I_{13/2}$ energy level - after absorption of a photon from the excitation - can be calculated by

$$P_{hop} = \frac{\Gamma_{ET,DD}}{\Gamma_{ET,DD} + \Gamma_{SPE} + \Gamma_{ET,DA}} . \quad (7)$$

with the rate $\Gamma_{ET,DD}$ for donor-donor energy transfer ($^4I_{13/2}, ^4I_{15/2}) \rightarrow (^4I_{15/2}, ^4I_{13/2})$, the rate $\Gamma_{ET,DA}$ for energy transfer upconversion by donor-acceptor energy transfer ($^4I_{13/2}, ^4I_{13/2}) \rightarrow (^4I_{15/2}, ^4I_{9/2})$, and the rate $\Gamma_{SPE}$ for spontaneous emission. These rates are given by

$$\Gamma_{ET,DD} = \frac{C_{DD}}{d^6} N_2 N_1 \quad (8)$$

$$\Gamma_{ET,DA} = \frac{C_{DA}}{d^6} N_2 N_2 \quad (9)$$

$$\Gamma_{SPE} = \frac{1}{\tau_{R,21}} N_2 = A_{21} N_2 \quad (10)$$

with $N_1$ being the relative occupation of the ground state ($^4I_{15/2}$), $N_2$ the relative occupation of the first excited state ($^4I_{13/2}$), $\tau_{R,21}$ the radiative lifetime of the first excited state, and $A_{21}$ the corresponding Einstein coefficient for spontaneous emission. Due to the low phonon energy in $\beta$–$NaYF_4$ multi-phonon relaxation from $^4I_{15/2} \rightarrow {}^4I_{13/2}$ is negligible.

The migration probability $P_{migrate}$ describes the probability that the excitation energy from a donor still resides on a donor after $n_{hop}$ hopping processes and can be derived by

$$P_{migrate} = (P_{hop})^{n_{hop}} = \left(\frac{\Gamma_{ET,DD}}{\Gamma_{ET,DD} + \Gamma_{SPE} + \Gamma_{ET,DA}}\right)^{n_{hop}} . \quad (11)$$

The number of hopping processes can then be determined by

$$n_{hop} = \frac{\ln(P_{migrate})}{\ln(P_{hop})} . \quad (12)$$

To calculate $n_{hop}$ the parameters $N_1$ and $N_2$ have to be known. Both of these parameters depend on the irradiance of the excitation and typically no analytical solution can be found for the real upconverter system, which makes rate equation model with numerical solutions necessary.[19] However, we are only interested in an meaningful estimation of $n_{hop}$. Therefore, we estimate the $N_1$ and $N_2$ from a 2 energy level system, which only considers absorption and spontaneous emission. The rate equation for $N_2$ in the steady state is then

$$\frac{dN_2}{dt} = 0 = -A_{21} N_2(t) + \int B_{12}(\omega) u(\omega) d\omega \, N_1(t) . \quad (13)$$

For low irradiances as used in this work $N_1 \gg N_2$. In such a situation, $N_2$ and $N_1$ can be approximated by

$$N_2 = \frac{\int B_{12}(\omega) u(\omega) d\omega}{A_{21}} = \frac{B_{12} \int g_{12}(\omega) u(\omega) d\omega}{A_{21}}$$

$$= \frac{B_{12} \int g_{12}(\omega) u(\omega) d\omega}{\frac{\hbar \omega_{laser}^3 n^3}{\pi^2 c^3} B_{12} \frac{g_2}{g_1}} = \frac{\int g_{12}(\omega) u(\omega) d\omega}{\frac{\hbar \omega_{laser}^3 n^3}{\pi^2 c^3} \frac{g_2}{g_1}} \quad (14)$$

$$N_1 = 1 - N_2, \qquad (15)$$

with $B_{12}$ being the Einstein coefficient for absorption, $g_{12}(\omega)$ the lineshape function of the corresponding transition, which is here the normalized ground state absorption spectrum ($^4I_{15/2} \to {}^4I_{13/2}$) taken from Ref.19, $u(\omega)$ the spectral energy density of the excitation, $n$ the refractive index, $c$ the speed of light in vacuum, $\omega_{laser}$ the angular frequency of the laser excitation, and $g_2$ and $g_1$ the degeneracies of the energy levels 1 ($^4I_{15/2}$) and 2 ($^4I_{13/2}$). In our case of $\beta$-NaYF$_4$ Er$^{3+}$, the overlap integral for absorption was determined by

$$\int g_{12}(\omega)u(\omega)d\omega = \int g_{12}(\omega)\frac{n}{c}I f_{exc}(\omega)d\omega \qquad (16)$$
$$= \frac{n}{c}I \int g_{12}(\omega) f_{exc}(\omega)d\omega = 1.7\times 10^{-14}\frac{n}{c}I$$

using the normalized excitation spectrum $f_{exc}(\omega)$, which is the laser spectrum (delta function with center wavelength at 1523 nm), and the irradiance of the laser excitation $I$ at 1523 nm. The values for $N_2$ are less than 5% for irradiances below 100 W/cm$^2$.

The hopping of the donor excitation energy in the crystal lattice can be described in a 3D random walk like manner. The hopping distance $d$ can be approximated by the average Er-Er distance, which can be calculated from the lattice parameters $a$ and $c$ along with the Er$^{3+}$ doping level $x$ by

$$d_{\beta-NaYF_4} = \left(\frac{a^2 c \sqrt{3}/2}{1.5 x}\right)^{1/3}. \qquad (17)$$

In a 3D random walk, the average migration distance $R_{migrate}$, which the donor excitation energy travels from the origin – the first excited ion - after $n_{hop}$ hopping steps, is determined by

$$R_{migrate} = \sqrt{3 n_{hop}} d. \qquad (18)$$

Combining the migration rate equation model with a random walk model allows the calculation of the average energy migration (EM) path length denoted as $R_{migrate}$. The likelihood of the migration path length $R_{migrate}$ as a function of the irradiance is shown in Figure 2b. Here, we used the dataset reported in Ref. 45. for the material system LiYF$_4$: Er$^{3+}$, which is very similar to NaYF$_4$: Er$^{3+}$. The determined $R_{migrate}$ values are in good agreement with estimations of several nanometers per millisecond of the energy donor lifetime for a Eu$^{3+}$ material.[43] The average $R_{migrate}$ decrease from around 17 nm to around 3.2 nm as the irradiance increases from a few W/cm$^2$ to 150 W/cm$^2$, which is caused by a higher energy transfer upconversion probability ($P_{ET,DA}$) due to the higher population density of excited states. In other words, it is more likely that neighboring Er$^{3+}$ can be found in the first excited state and upconversion occurs (power law). As a result, for higher irradiances the likelihood of migration of excited states to the surface of the UCNCs is reduced, where the excitation energy might be quenched.

## 4. Experimental Results and Discussion

UCNCs with $\beta$-NaYF$_4$: Er$^{3+}$ cores and inert $\beta$-NaLuF$_4$ were synthesized as described in the experimental section. TEM images were used to determine the diameters of the core and the core-shell nanocrystals, as shown for CS2 in Figure 3a,b. The CS2 nanocrystals are monodisperse with a standard deviation of the diameter of less than 6% (Figure 3c). The optically active $\beta$-NaYF$_4$: 28% Er$^{3+}$ cores feature diameters of 19.2±1.1 nm. A very isotropic and thick inert shell could be grown around the core with a thickness of approximately 10 nm owing to the tensile strain between the core material ($\beta$-NaYF$_4$: 28% Er$^{3+}$) and the shell material ($\beta$-NaLuF$_4$).[26]

The ionic radii of the lanthanide (Ln) series decrease for larger element numbers, which is called the lanthanide contraction. Consequently, the lattice parameters of $\beta$-NaLnF$_4$ decrease for heavier Ln ions in the crystals. This was reported by Thoma et al. for the Ln series as well as $\beta$-NaYF$_4$, which has lattice parameters ranging between the ones of $\beta$-NaHoF$_4$ and $\beta$-NaErF$_4$.[50] Johnson and van Veggel observed in their comprehensive study along the Ln series a more anisotropic shell growth, when the lattice parameters of the shell were larger than the parameters of the core, which results in compressive strain in the epitaxial layer. On the other hand, they determined a more isotropic shell growth when the lattice parameters of the shell were smaller than the parameters of the core material, which results in tensile strain in the epitaxial layer.[26] For this reason, we chose a core-shell system with tensile strain in the epitaxial layer and β-NaLuF$_4$ as the inert shell material.

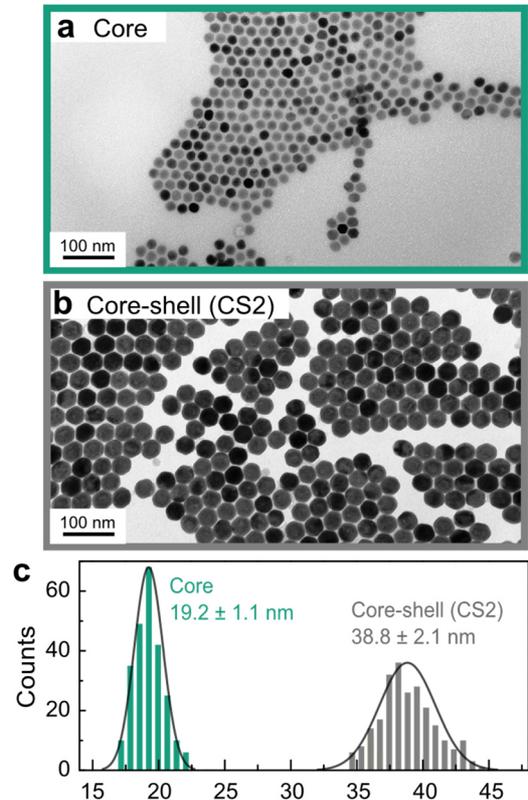

**Figure 3** (**a,b**) TEM images of the active core $\beta$-NaYF$_4$: 28% Er$^{3+}$ UCNCs and the CS2 UCNCs with very thick and isotropic $\beta$-NaLuF$_4$ shell. The tensile strain of the shell enabled the growth of a very isotropic and thick shell. (**c**) Measured size distribution of the core and CS2 UCNCs.

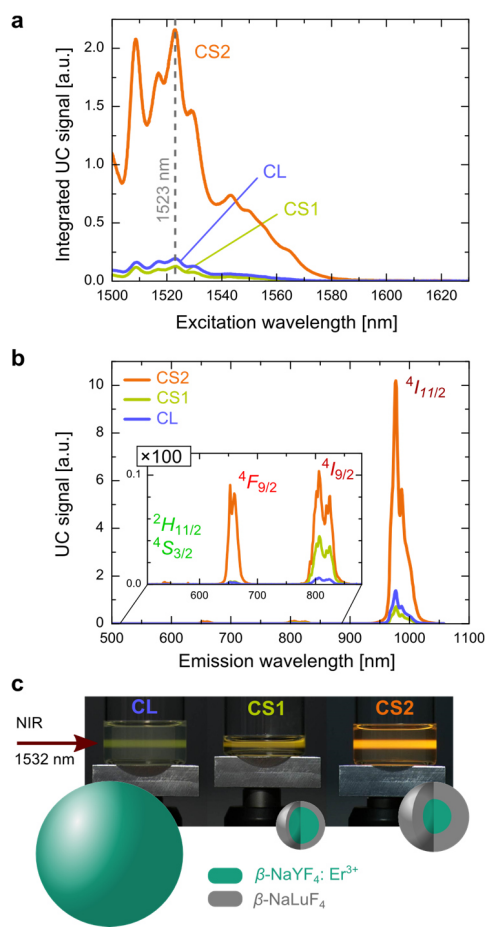

**Figure 4.** Spectroscopic properties of the UCNCs in chloroform. (**a**) The excitation profiles of all the different UCNCs show a peak of the integrated upconversion signal at 1523 nm. (**b**) Upconversion emission spectrum for an excitation wavelength of 1523 nm with an irradiance of 0.43 ± 0.03 W/cm². The dominant upconversion emission originates from a 2-photon process from the transition $^4I_{11/2} \rightarrow {^4I_{15/2}}$ with a center wavelength of 980 nm. (**c**) The photographs of the different UCNCs in CHCl$_3$ were taken with the same capture settings and with ambient light. The nanocrystal sketches illustrate the different proportions of the investigated nanocrystals true to scale. Strong visible upconversion was observed originating from at least 3-photon processes.

In our study, we compare the core-shell UCNCs with very thick and isotropic shell (CS2) with two reference materials: "large" UCNCs (CL) with a diameter of 84.0±32.0 nm and commonly used core-shell UCNCs (CS1) with a core diameter of 20.0±1.6 nm and an approximately 3.6 nm thick and isotropic β-NaLuF$_4$ shell. The concentrations of the UCNCs in CHCl$_3$ were chosen to achieve a similar concentration of Er$^{3+}$ ions per volume of the solution (Table 2). The excitation and emission spectra of the UCNCs in CHCl$_3$ are shown in Figure 4a,b.

The excitation profiles were measured using an irradiance of 0.38±0.03 W/cm², which is within the applicable range for photovoltaics (solar concentration factor of ~160×; 0.0024 W/cm² over the range from 1480 - 1580 nm for 1 sun). The integrated upconversion signal from the CS2 UCNCs is 17× higher than that of the reference materials. The upconversion emission spectra shown in Figure 4b were recorded using an excitation wavelength of 1523 nm and an irradiance of 0.43±0.03 W/cm². The UC emission from the transition $^4I_{11/2} \rightarrow {^4I_{15/2}}$ with a center wavelength of 980 nm (NIR) is the dominant one with around 98% of total emission for the CS2 UCNCs. In addition to the enhancement of the $^4I_{11/2} \rightarrow {^4I_{15/2}}$ transition, the CS2 UCNCs show a strong enhancement of the $^4F_{9/2} \rightarrow {^4I_{15/2}}$ transition, which is centered around 655 nm (red). The ratio of this transition on the total UC emission is enhanced by a factor of more than 2 compared to the reference core-shell (CS1) and around 5 with respect to the "large" nanocrystals (CL). This enhancement is responsible for the yellowish color impression visible in the photographs of the CS2 UCNCs (Figure 4c).

**Table 2.** Material parameters of the different upconverter nanocrystal (UCNC) samples. The internal UCQY values were determined for an excitation wavelength of 1523 nm using an irradiance of 0.43±0.03 W/cm².

|  | "Large" (CL) | Core-shell (CS1) | Core-thick shell (CS2) |
|---|---|---|---|
| Er$^{3+}$ doping [%] | 10 | 28 | 28 |
| Diameter core [nm] | 84.0±32. | 20.0±1.6 | 19.2±1.1 |
| Thickness of the shell [nm] | 0 | 3.6 | 9.8 |
| Initial concentration in solution [UCNCs/ml] | 1.7 × 10$^{13}$ | 8.8 × 10$^{14}$ | 5.8 × 10$^{14}$ |
| Er$^{3+}$ per nanocrystals | 418,000 | 15,700 | 14,300 |
| Er$^{3+}$ concentration [Er$^{3+}$/ml] | 7.1 × 10$^{18}$ | 1.4 × 10$^{19}$ | 8.4 × 10$^{18}$ |
| Internal UCQY in CHCl$_3$ [%] | 0.13±0.03 | 0.04(4)±0.01 | 0.71±0.08 |
| Internal UCQY in PMMA [%] | 0.24±0.07 | 0.38±0.13 | 2.01±0.19 |

We then determined (i) the external UCQY, defined as the ratio of emitted upconverted photons to the incident photons, and (ii) the internal UCQY, defined as the ratio of emitted upconverted photons to absorbed photons. Three different concentrations of the CS2 UCNCs in CHCl$_3$ were investigated. The external UCQY increases for higher concentrations due to increased absorption (Figure 5a). The internal UCQY is independent from the concentration of the UCNCs in the solvent, as expected (Figure 5b). The weighted average values of the three measurements with different nanocrystal concentrations were calculated. The internal UCQY reaches 0.71±0.08 % for the CS2 UCNCs using an irradiance of 0.43±0.03 W/cm² and an excitation wavelength of 1523 nm. No saturation of the UCQY was observed. Consequently, higher UCQY values are expected for higher irradiances (Supporting Information, Figure S8).

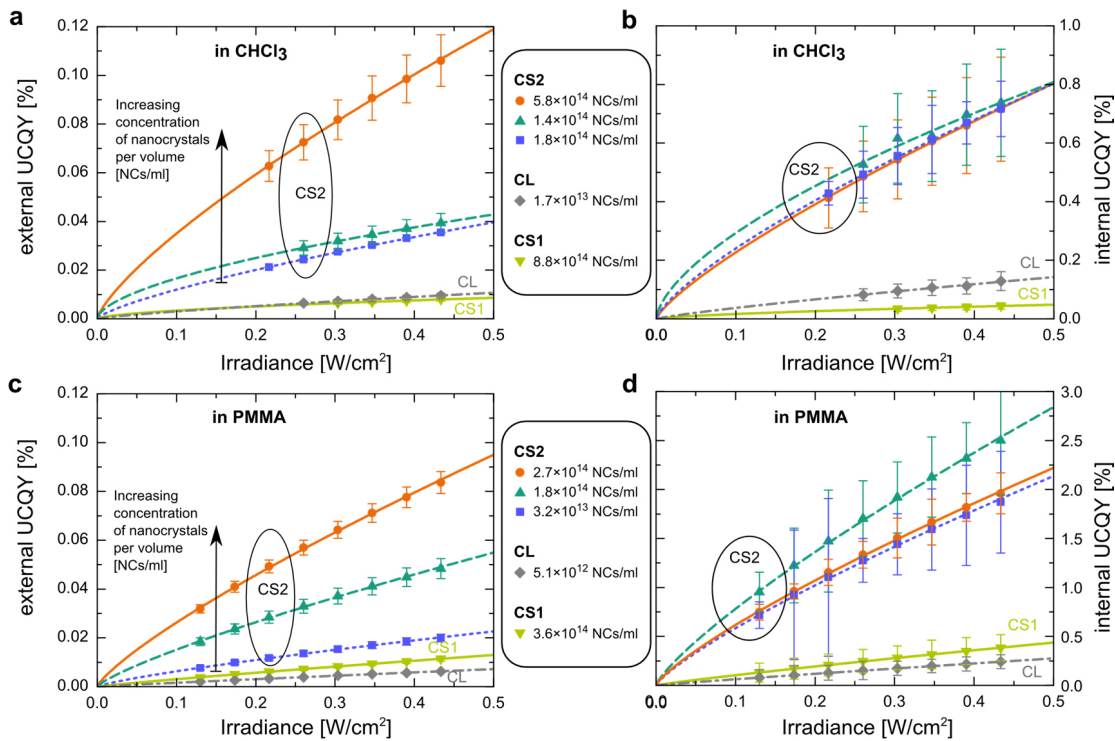

**Figure 5.** Upconversion quantum yield (UCQY) analysis. (**a**) External and (**b**) internal UCQY of the UCNCs in the solvent $CHCl_3$. The external UCQY increases for higher nanocrystal concentrations due to increased absorption, while the internal UCQY is independent of the nanocrystal concentration. (**c**) External and (**d**) internal UCQY of UCNCs-PMMA composites. The CS2 UCNCs in embedded in PMMA are around 2.8× more efficient than the same nanocrystals dispersed in $CHCl_3$.

Typically, irradiances of more than 3 orders of magnitude higher than the ones used in this work are applied in UCQY measurements of UCNCs.[28, 49, 50] Using the normalized UCQY of 0.017±0.002 cm²/W, which is a commonly used figure of merit as discussed above and in Ref. 22, the CS2 UCNCs are more than 43× more efficient than the best $Er^{3+}$-doped nanomaterials reported to date in Ref. 31, demonstrating their high potential for successful integration in many applications.

To form upconverter solar cell devices, which exploit sub-bandgap photons from the solar radiation, the upconverter needs to be applied on the rear side of the solar cells. For this purpose, we embedded the different UCNCs with different concentrations into the polymer PMMA to form solid, stable, and transparent UCNCs-PMMA composites (Supporting Information Figure S1). In PMMA, the internal UCQY of all UCNCs is higher compared to the UCNCs in $CHCl_3$ (Figure 5d). The internal UCQY of the CS2 UCNCs-PMMA composites reaches 2.01±0.19 % using an irradiance of 0.43±0.03 W/cm² at 1523 nm. This value is 120× higher than the previous best value we found in the literature for solely $Er^{3+}$-doped core-shell nanocrystals with respect to the irradiance with a normalized UCQY of 0.047±0.005 cm²/W.[31] Even compared to co-doped systems, which operate at different excitation wavelengths and are commonly known for their superior performance, the CS2 UCNCs are ≥260× more efficient than nanocrystals doped with $Er^{3+}$ and $Yb^{3+}$ (1.78×10⁻⁴ cm²/W)[51] and ≥100× more efficient than nanocrystals doped with $Yb^{3+}$ and $Tm^{3+}$ (4.49×10⁻⁴ cm²/W).[52]

In composite form, the upconverting nanocrystal can directly be used in photovoltaics and therefore can be compared to bulk materials. The internal UCQY value of the CS2 UCNCs-PMMA composites is already 1/6 of 12.0±1.0 %, which is the highest value reported for the corresponding bulk material at nearly the same irradiance of 0.40±0.02 W/cm² and therefore the most meaningful comparison.[30] In addition, the bulk material in Ref. 30 shows strong saturation of the UCQY at this irradiance level, whereas no saturation of the UCQY is observed for the UCNCs. In consequence, the gap of the internal UCQY between the CS2 UCNCs and the best bulk materials is decreasing at higher irradiances, which is supported by extrapolations of the internal UCQY to higher irradiances (see Supporting Information, Figure S8, Table S5). Furthermore, the normalized internal UCQY value of the UCNCs-PMMA composites exceeds values reported for other bulk materials.

As a proof of principle, the CS2 UCNCs-PMMA composites were applied to the rear side of a calibrated bifacial crystalline silicon solar cell (Figure 6a). External quantum efficiency (EQE) due to upconversion of sub-bandgap photons of 0.068±0.002 % was achieved for an irradiance of 0.45±0.03 W/cm² at 1523 nm, demonstrating the successful realization of a nanocrystal upconverter solar cell device. This EQE can be enhanced considerably when the absorption by the upconverter sample is increased, for example, in combination with photonic structures or by simple increasing the amount of nanocrystals in the composite. Here, the UCNCs took only 0.83% of the complete sample's volume for a concentration of $2.4×10^{14}$ NCs/ml. If the absorptance of the UCNCs-PMMA composite could be scaled up to to the bulk value of 66% [14, 30], for example by increasing the concentration of NCs/ml, we estimated an $EQE_{UC}$ of 1.18 % for an UCNCs photovoltaic device. The corresponding bulk $EQE_{UC}$ value is around 5.7%. This means that the UCNCs in a photovoltaic device would perform only 1/5 less efficient than the values achieved with the very best bulk materials available for silicon solar cells, which are $\beta$-NaYF$_4$: 25% Er$^{3+}$ microcrystals embedded in a polymer.[22, 32, 55, 56] Other material systems, such as Er$^{3+}$-doped Gd$_2$O$_2$S microcrystals or BaY$_2$F$_8$ single crystals have also shown high $EQE_{UC}$ values, in some cases even exceeding the upconverter solar cell device performance of $\beta$-NaYF$_4$: Er$^{3+}$.[16, 22, 55] In addition, the scaled $EQE_{UC}$ of the UCNCs-PMMA composite is in very good aggreement with our optically determined internal UCQY.

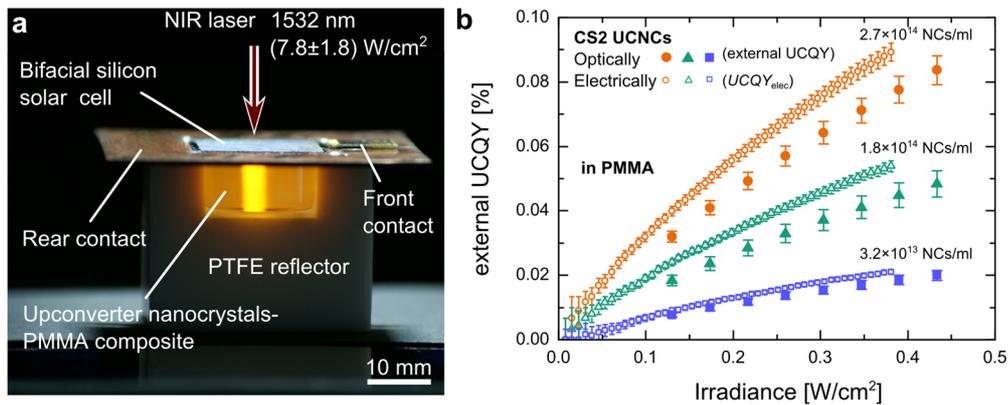

**Figure 6.** Upconverter solar cell device using the CS2 UCNCs-PMMA composites. (a) Cross-section photograph of an upconverter solar cell device illuminated with a high power NIR laser, which is transmitted through the bifacial silicon solar cell. (b) The external UCQY as calculated from the electrical measurement of upconverter solar cell devices is in good agreement with the results from the optical measurement using photoluminescence spectroscopy. Photon recycling due to the highly reflecting PTFE reflector results in the slightly higher external UCQY in the electrical measurements.

For a direct comparison of the EQE to the optically determined external UCQY, one has to consider the incomplete transmittance of the excitation light through the solar cell and the efficiency by which the upconverted photons are utilized at the rear side of the solar cell.[15, 16] This quantity, denoted as $UCQY_{elec}$, is in good agreement with the optical measurements (Figure 6b). The trend of higher $UCQY_{elec}$ values can be attributed to the highly reflecting polytetrafluoroethylene (PTFE) mounting, which leads to an increased photon recycling in the electrical measurements, for example, by a longer effective path length of the excitation light in the upconverter sample.

To elucidate as to why the CS2 UCNCs show such an exceptional performance, we performed further analyses to answer two questions: (i) why is the shell especially beneficial when it is isotropic and thick (~10 nm), and (ii) why did the embedding of the nanocrystals in the polymer matrix increase the internal UCQY even further?

Static surface quenching processes are a result of crystal defects at the surface and energy transfer processes from the rare-earth ions to the ligands, the organic molecules of the solvent, or other quenchers (e.g. water) near or on the surface. For our samples, we did not measure any significant quenching effect due to residual water molecules in the solution of CHCl$_3$. Furthermore, we excluded vibrational modes of C-H molecules in the solvent CHCl$_3$ to be responsible for the quenching of the upconversion luminescence in a comparative study using *d*-chloroform (Supporting Information, Figures S5-S6). Hence, we conclude that surface defects as well as energy transfer processes to the ligands, hydroxyl groups or also other quenchers at the surface are relevant for the surface quenching, but effectively suppressed in sample CS2. The strong increase in the internal UCQY due to the embedding in PMMA would also fit into this picture, as the PMMA immobilizes the ligands and influences the surface conditions of the nanoparticles, which may act as an additional surface passivation.

As a next step, we discuss how these effects depend on the size of the nanocrystal and the thickness of the shell. The critical distance for energy transfer from rare-earth ions in nanocrystals to organic compounds close to the surface was reported to be 7 nm.[53] Hence, a passivating shell should be thicker than this to reduce effectively the quenching via resonant energy transfer. The shell should also be isotropic, as the thinnest part of the shell limits the overall passivation of the active core. The second relevant quenching pathway is determined by the surface of the nanocrystal itself, which may have a large number of defects owing to the open inorganic bonds. The surface defects can be healed by growing an epitaxial shell, which serves as a surface passivation. At the new outer emerging surface new defects arise but now at a larger distance from the optically active core and therefore with reduced surface-related quenching. Consequently, there will be an interrelation between the effective passivation quality of the surface defects and the thickness of the shell.

For both above described effects it is important to know which fraction of the rare-earth ions in the active volume is close enough to the surface to be affected by surface quenching. Here, also energy migration (EM) of the excitation energy has to be taken into account. EM can be understood as a hopping process of the excitation energy from one ion to another ion in a random walk manner as described in Section 3.[54]

In our study, the most conservative $R_{migrate}$ value is 3.2 nm for an irradiance of 150 W/cm$^2$, see Figure 2. For a spherical particle with a diameter of 20 nm, about 70% of the rare-earth ions are found in a shell volume within a distance of 3.2 nm from the surface, compared to 21% for a nanocrystal with a diameter of 84 nm. The excitation energy from these ions can reach the surface and be quenched. For an irradiance of 0.5 W/cm$^2$ a $R_{migrate}$ value of 16 nm was calculated, which means that statistically the energy of every excited ion can migrate to the surface for 20 nm diameter particles. For particles with diameters of 84 nm, 73% can reach the surface. Although, one has to keep in mind that the quenching rate at the surface has a finite value, which means that not all the excitation energy that migrates to the surface is quenched, this analysis explains why surface quenching is such an important effect. Furthermore, we revealed that the energy migration path length decreases with increasing irradiance. This indicates the need of high irradiances for relevant upconversion luminescence without sufficient surface passivation of the optically active core.

This whole set of particle-size dependent effects comes on top of the fact that embedding of the UCNCs into PMMA may result in further reduced surface quenching, for example by immobilizing of ligands or hydroxyl groups on the surface of the nanocrystals due the rigid polymer matrix. The ligands and other quenchers can no longer move freely, which may change the vibrational energies and could lead to less efficient surface quenching. This could explain why the increase of the internal UCQY due to the embedding of nanocrystals in PMMA is the lowest for the CL ("large") UCNCs, with the lowest surface-to-volume ratio and the highest for the smaller CS1 UCNCs with thinner shell and highest surface-to-volume ratio. We determined the ratio of the internal UCQY of the UCNCs embedded in PMMA to solution in CHCl$_3$. The averaged ratio for all considered irradiances is shown in Figure 7. An enhancement factor of 1.8 was obtained for the CL UCNCs, whereas enhancement factor of 8.6 and 2.8 were determined for the much smaller CS1 UCNCs and the CS2 UCNCs, respectively.

## 5. Conclusions

Using a rate equation model to describe the migration of excitation energy in the nanocrystal, we determined an average migration path length of several nanometers. This means that a large fraction of the excitation energy also migrates to the surface, where the excited states may be quenched. Therefore, a good surface passivation is not only important for the excited lanthanide ions directly at the surface but even more for the complete lanthanide nanocrystal. Additionally, we found that the average migration energy strongly depends on the irradiance and decreases as the irradiance increases.

We synthesized upconverting nanocrystals (UCNCs) and used $\beta$-NaLuF$_4$ with tensile strain as the inert shell material to grow very isotropic and thick shells around the $\beta$-NaYF$_4$: Er$^{3+}$ cores. In our comprehensive experimental upconversion quantum yield (UCQY) study, we show that UCNCs with thick $\beta$-NaLuF$_4$ show an exceptional good performance. Our findings indicate that an inert shell thickness of around 10 nm is necessary to protect the optically active core sufficiently from surface quenching processes. In solution, an internal UCQY of 0.71±0.08% was determined for an irradiance at 1523 nm of only 0.43±0.03 W/cm$^2$, relevant for photovoltaics. Furthermore, we show that embedding of UCNCs in a polymer enhances the UCQY considerably by a factor of 2.8 to a UCQY of 2.0±0.2 % using a low irradiance of 0.43±0.03 W/cm$^2$ at 1523 nm. As a result, the CS2 upconverting nanocrystals with around 10 nm thick and isotropic shell are 120× more efficient than best comparable nanomaterial found in the literature with respect to the irradiance of the excitation. Additionally, we report internal UCQY values of the CS2 upconverter nanocrystals reaching 1/6 of the highest reported value for $\beta$-NaYF$_4$: Er$^{3+}$ bulk material, which was reported to be in the range of 1/750 before this paper. Furthermore, saturation of the internal UCQY was not yet observed and therefore higher internal UCQY at higher irradiances are expected.

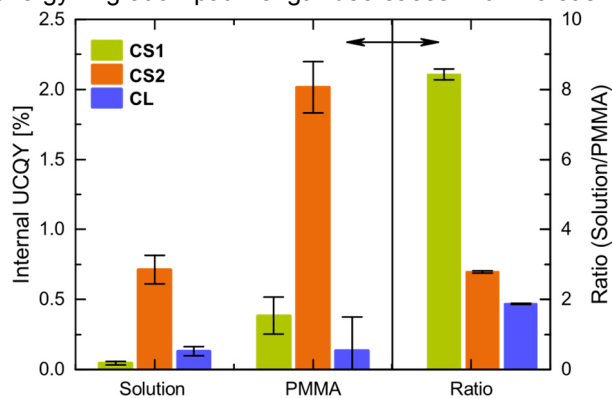

**Figure 7.** Internal UCQY in solution of CHCl$_3$ and embedded in PMMA for an excitation wavelength of 1523 nm and an irradiance of 0.43±0.03 W/cm$^2$. Here, the ratios of the internal UCQY from UCNCs in solution and in PMMA were averaged over all considered irradiances and the standard derivation was calculated. The enhancement of the internal UCQY due to embedding in PMMA is stronger for smaller UCNCs and for thinner shells indicating dynamic quenching to be less significant in larger particles and more efficient surface passivation.

As a proof of concept, we applied the upconverter nanocrystal-PMMA composites on the rear side of a bifacial crystalline silicon solar cell. The results of the upconverter solar cell device measurements validated our optical findings and the high internal UCQY values at low irradiance determined for the upconverting nanocrystals with very thick and isotropic shell.

The very thick and isotropic shell concept is transferable to other material systems. For example, this concept is promising for highly efficient and bright nanoprobes in life sciences, and to produce alternative phosphors for white LEDs, where a high quantum yield is essential for energy efficient lighting. A great benefit of nanocrystals is the fact that scattering of light is typically negligible as compared to commonly used microcrystalline phosphors, which might give further flexibility in the spatial radiation behavior of the LED as well as in the color tuning.

**Acknowledgements**

Stefan Fischer gratefully acknowledges the scholarship support from the Deutsche Bundesstiftung Umwelt (DBU). The research leading to these results has received funding from the European Community's Seventh Framework Programme (FP7/2007-2013) under grant agreement n°[246200] within the Nanospec project. The Natural Science and Engineering Council (NSERC) of Canada is gratefully acknowledged for financial support.

# Supporting Information

# Upconverting Core-Shell Nanocrystals with High Quantum Yield under Low Irradiance: On the Role of Isotropic and Thick Shells

Stefan Fischer[1], Noah J.J. Johnson[2], Jothirmayanantham Pichaandi[2], Jan Christoph Goldschmidt[1], and Frank C.J.M van Veggel[2]

[1]*Fraunhofer Institute for Solar Energy Systems, Heidenhofstraße 2, 79110 Freiburg, Germany*
[2]*Department of Chemistry, University of Victoria, P.O. Box 3065, Victoria, British Columbia, Canada V8W 3V6*

## Contents

# 1. Overview over Materials and Results

**Table S1**. The material properties of the investigated nanocrystals are summarized in this table. For core-"giant" shell nanocrystals, the molar ratio of the shell material to the core material is approximately 6.8, assuming a reasonable chemical yield of 80% [1] for the $\beta$-NaYF$_4$: Er$^{3+}$ core reaction. This also matches well to the core-to-shell volume ratio, which was calculated from the diameters determined from the transmission electron microscopy (TEM) images assuming spherical particles. The values of the concentrations in nanocrystals per ml and the Er$^{3+}$ concentrations in Er$^{3+}$ per ml refer to the initial solution in CHCl$_3$.

|  | "large" | Reference | Core-shell | Core-"giant" shell |
|---|---|---|---|---|
| Er$^{3+}$ doping [%] | 10 | 0 | 28 | 28 |
| Diameter core [nm] | 84.0±32.0 | 16.0±1.8 | 20.0±1.6 | 19.2±1.1 |
| Diameter core-shell [nm] |  | 20.5±1.4 | 27.4±2.4 | 38.8±2.1 |
| Thickness shell [nm] |  | 2.2 | 3.6 | 9.8 |
| Volume ratio shell-to-core |  | 1.1 | 1.5 | 7.3 |
| Concentration [UCNCs/ml] | 1.7 × 10$^{13}$ | 3.7 × 10$^{14}$ | 8.8 × 10$^{14}$ | 5.8 × 10$^{14}$ |
| Er$^{3+}$ per nanocrystals | 418 000 |  | 15 700 | 14 300 |
| Er$^{3+}$ concentration [Er$^{3+}$/ml] | 7.1 × 10$^{18}$ |  | 1.4 × 10$^{19}$ | 8.4 × 10$^{18}$ |

**Table S2. Upconverter nanocrystals samples in solution using the solvent CHCl$_3$.** The UCQY values were determined for an excitation wavelength of 1523 nm using an irradiance of 0.43±0.03 W/m$^2$. The weighted average of the internal UCQY of the core-"giant" shell UCNCs is 0.713±0.076 %.

| Sample | Concentration [UCNCs/ml] | Absorptance [%] | UCQY [%] external | UCQY [%] internal |
|---|---|---|---|---|
| NC1 (core-"giant" shell) | 5.8 × 10$^{14}$ | 15.0±3.3 | 0.106±0.011 | 0.716±0.178 |
| NC2 (core-"giant" shell) | 1.4 × 10$^{14}$ | 5.4±1.2 | 0.039±0.004 | 0.738±0.183 |
| NC3 (core-"giant" shell) | 1.8 × 10$^{14}$ | 4.9±1.1 | 0.035±0.001 | 0.716±0.095 |
| NC4 (core-shell) | 8.8 × 10$^{14}$ | 18.2±4.0 | 0.008±0.001 | 0.044±0.011 |
| NC5 ("large") | 1.7 × 10$^{13}$ | 7.6±1.7 | 0.009±0.001 | 0.129±0.032 |

**Table S3. Upconverter nanocrystals-PMMA composite samples.** The UCQY values were determined for a wavelength of 1523 nm using an irradiance of 0.43±0.03 W/m$^2$. The weighted average of the internal UCQY of the core-"giant" shell UCNCs is 2.014±0.190 %. The absorptance values in brackets were determined using the method with the spectrophotometer, as described in the supplemental text above.

| Sample | Concentration [UCNCs/ml] | Absorptance [%] | UCQY [%] external | UCQY [%] internal |
|---|---|---|---|---|
| P1 (core-"giant" shell) | 2.7 × 10$^{14}$ | 4.3±0.5 (3.8) | 0.084±0.005 | 1.960±0.210 |
| P2 (core-"giant" shell) | 1.8 × 10$^{14}$ | 1.9±0.4 (3.2) | 0.048±0.004 | 2.502±0.560 |
| P3 (core-"giant" shell) | 3.2 × 10$^{13}$ | 1.1±0.4 (1.6) | 0.020±0.002 | 1.871±0.732 |
| P4 (core-shell) | 3.6 × 10$^{14}$ | 3.0±1.0 (3.9) | 0.011±0.001 | 0.384±0.134 |
| P5 ("large") | 5.1 × 10$^{12}$ | 2.6±0.7 (2.7) | 0.006±0.001 | 0.241±0.070 |

## 2. Embedding of Nanocrystals in PMMA

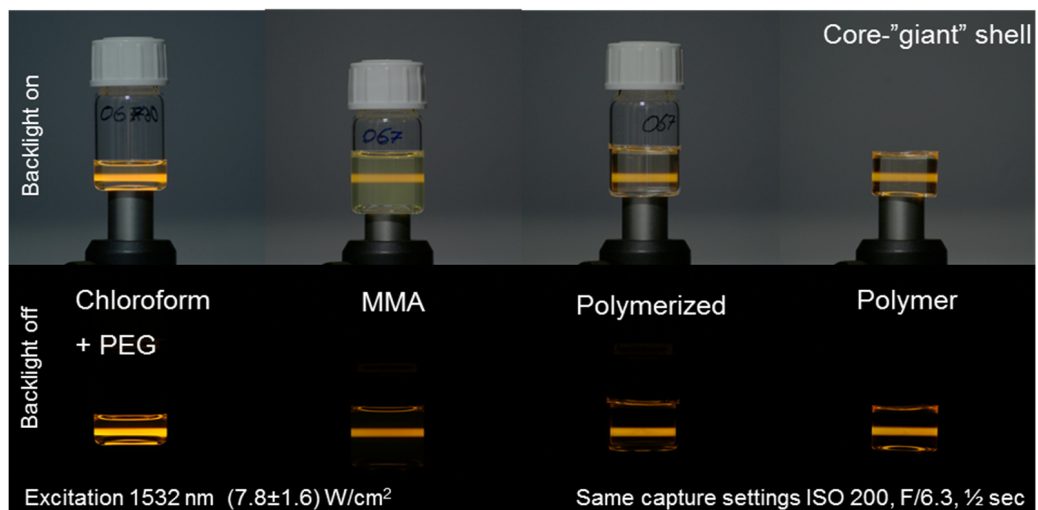

**Figure S1.** Photographs taken at each step of the fabrication of the nanocrystals-PMMA composites show the strong visible upconversion luminescence under excitation with a NIR laser diode (4PN-1532-6-95, Laser Components) with a wavelength of 1532 nm and an irradiance of 7.8±1.8 W/cm². The visible luminescence is a result of upconversion processes involving at least 3 photons. The pictures were taken with and without backlight illumination and with the same capture settings for all pictures. Due to the dilution of the UCNCs in the volume, the UCNCs-PMMA composite samples appear darker than the samples in chloroform (CHCl₃).

## 3. Methods and Setup to determine the UCQY

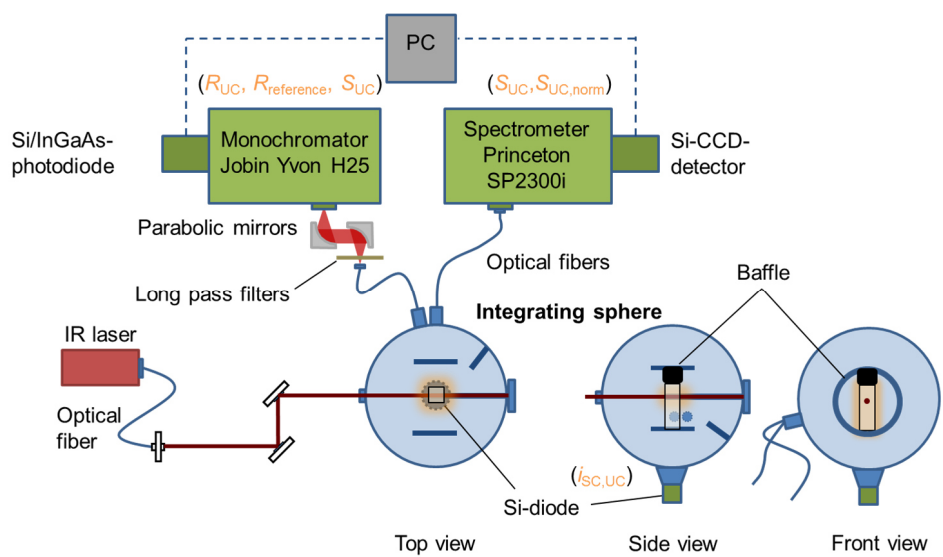

**Figure S2.** Schematic of the experimental setup to determine the external and internal upconversion quantum yield (UCQY) under monochromatic laser excitation via an integrating sphere. The cuvettes filled with the UCNCs in solution or the UCNCs-PMMA composite samples are placed in the center of the integrating sphere. Optical fibers guide the light from the integrating sphere to a monochromator and a spectrometer in order to measure the laser spectrum and the upconversion luminescence. The integrated upconversion luminescence can also be measured with an adapted and calibrated silicon (Si) photodiode. Several baffles prevent the reflected laser light or the emitted light from the upconverter samples to be detected directly.

As discussed in the experimental section of the paper, we distinguished between two definitions for the UCQY: the *external UCQY*

$$eUCQY = \frac{\text{upconverted photon flux}}{\text{incident photon flux}} = \frac{\phi_{UC}}{\phi_{in}}. \tag{S1}$$

and the *internal UCQY*

$$iUCQY = \frac{\text{upconverted photon flux}}{\text{absorbed photon flux}} = \frac{\phi_{UC}}{\phi_{abs}} = \frac{1}{A_{UC}} eUCQY \quad . \quad (S2)$$

The incident photon flux from the NIR laser $\phi_{in}$ was determined using a calibrated germanium detector, which was placed close to the sample position and illuminated with the laser beam. The incident photon flux $\phi_{in}$ was calculated as

$$\phi_{in} = \frac{i_{SC,exc} \int S_{exc}(\lambda)(1-R_p)(\lambda) d\lambda}{e \int S_{exc}(\lambda) EQE_{Ge}(\lambda) d\lambda} \quad (S3)$$

using the short-circuit current of a germanium detector $i_{SC,exc}$, the external quantum efficiency of the germanium detector $EQE_{Ge}$, the normalized excitation spectrum $S_{exc}$ from the NIR laser (delta function), the elementary charge $e$, and a reflection term $R_p$, which considers parasitic reflections of the excitation laser beam reducing the incident photon flux. The irradiance $I$ of the monochromatic laser excitation with wavelength $\lambda_{laser}$ can easily be calculated from $\phi_{in}$ and the laser beam area $A_{laser}$.

$$I(\lambda_{laser}) = \frac{\phi_{in}}{A_{laser}} \frac{hc}{\lambda_{laser}} \quad (S4)$$

In the experiments, the relative upconversion emission spectrum $S_{UC}$ was measured with the spectrometer. This spectrum was normalized to an integrated value of one $S_{UC,norm}$. In addition, the short-circuit current due to upconverted photons $i_{SC,UC}$ was measured using the silicon photodiode, which is attached directly to the integrating sphere. The background signal was subtracted so that only the upconverted photons contribute to the short-circuit current. Hence, the photon flux of upconverted photons $\phi_{UC}$ is given by

$$\phi_{UC} = \frac{1}{\beta} \frac{i_{SC,UC}}{e \int S_{UC,norm}(\lambda) EQE_{Si}(\lambda)(1-R_p)(\lambda) d\lambda} , \quad (S5)$$

with the collection efficiency of the integrating sphere $\beta$, the normalized upconversion emission spectrum $S_{UC,norm}$, and the external quantum efficiency of the Si photodiode $EQE_{Si}$. The overall error of the external UCQY is dominated by the uncertainty of the irradiance $I$ and the collection efficiency $\beta$. Typical systematic errors of the external UCQY are around 7.1%.

The signal of the NIR laser with the upconverter sample $R_{UC}$ and with the reference sample $R_{reference}$, showing no upconversion but the same scattering properties, were measured using the monochromator along with the InGaAs detector to determine the absorptance of the upconverter sample $A_{UC}$ at the laser wavelength. $A_{UC}$ was calculated by

$$A_{UC} = \frac{\int (R_{reference}(\lambda) - R_{UC}(\lambda)) d\lambda}{\int R_{reference}(\lambda) d\lambda} = 1 - \frac{\int R_{UC}(\lambda) d\lambda}{\int R_{reference}(\lambda) d\lambda} . \quad (S6)$$

The error of the internal UCQY strongly depends on the uncertainty (statistical error) of the absorptance $A_{UC}$. In this work, the errors are typically around 10% for samples showing high absorptance levels, but increase up to 25% for samples with lower absorptance.

Another method to determine optical quantum yields is more commonly used in the literature.[2,3] The measured signals of the upconverter sample and the reference are directly used to calculate the internal UCQY

$$iUCQY = \frac{\int S_{UC}(\lambda) d\lambda}{\int (R_{reference}(\lambda) - R_{UC}(\lambda)) d\lambda} . \quad (S7)$$

For this method the setup has to be calibrated on an absolute scale or the two detection setups have to be aligned exactly to each other. Therefore, we used solely the monochromator with the InGaAs detector, which covers the spectral range from the excitation to the upconversion emission. Therefore, no error-prone alignment between the two detection units – spectrometer and monochromator - is necessary. In the experiment, only the emission from the transition $^4I_{11/2} \rightarrow {}^4I_{15/2}$ at around 980 nm was detected. This $^4I_{11/2} \rightarrow {}^4I_{15/2}$ transition contributes with more than 98% on the total upconversion luminescence in the case of the core-"giant" shell UCNCs. Consequently, the absolute internal UCQY is well represented by the $^4I_{11/2} \rightarrow {}^4I_{15/2}$ transition. The main experimental challenge of this method is to measure the upconversion luminescence, which is very broad and weak, and the laser excitation, which is very narrow and strong, with the same setup settings.

We used both techniques (Equation S2 and S7) to determine the internal UCQY of the core-"giant" shell UCNCs. A very good agreement was found as shown in Figure S3a.

# 4. Performing of the UCQY Measurements

The external and internal UCQY of the nanocrystals were determined with the setup and methods introduced above.

**In solution:**

Samples were prepared with different concentrations of the nanocrystals in solutions with the solvent $CHCl_3$. These solutions were filled into cuvettes for the photoluminescence measurements. The cuvettes were introduced into the center of the integrating sphere. Three identical cuvettes were available. These cuvettes were each filled with different samples of the UCNCs in solution. The three cuvettes were measured alternately (measurement round). At least 10 measurement rounds were performed and the mean value of this measurement series was calculated.

The internal UCQY of the core-"giant" shell UCNCs was determined using two different methods, as discussed above. The results of the two methods agree very well, as shown in Figure S3a. Two measurement series for each characterization method were performed. The measurements of each series were performed with one week in between. An excellent agreement between the two methods could be achieved.

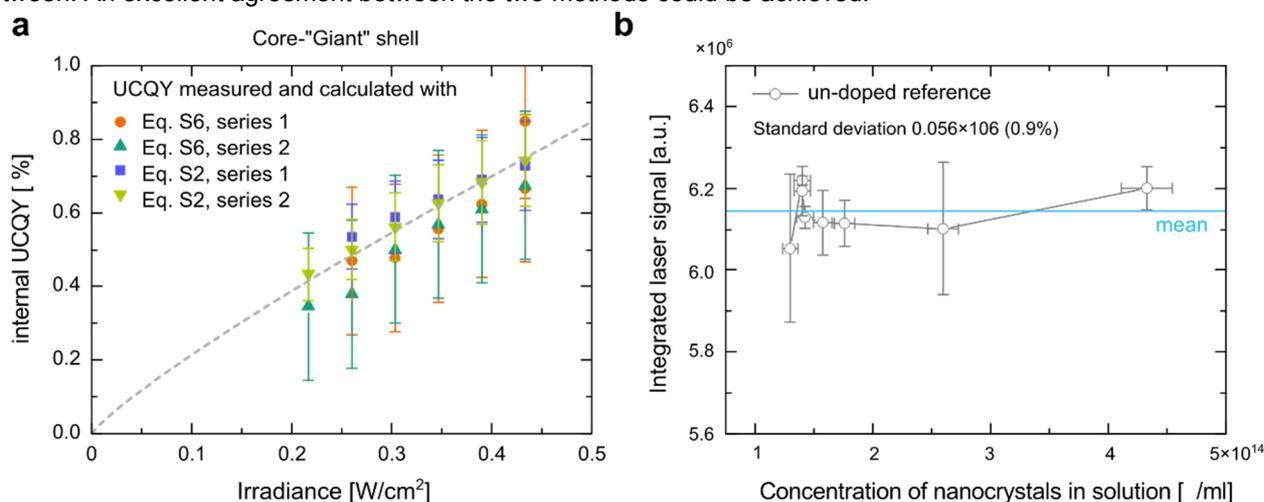

**Figure S3.** (**a**) The internal UCQY was measured for the same sample with the two different methods (Equation S2 and S7). Furthermore, with each method the internal UCQY was measured at different dates with one week in between. All measurements agree very well. (**b**) The integrated signal of the excitation laser was measured as a function of the concentration of the un-doped reference nanocrystals in the solution. For each sample, only 3 measurements were performed, which results in the fairly large errors and statistical fluctuations of the measured integrated laser signal in the range of less than 1.6%. No clear trend of the integrated laser signal with the concentration of the nanocrystals could be observed. Consequently, the same reference sample was used for all calculations of the absorptance of the UCNCs.

**In PMMA:**

The UCNCs-PMMA composites were introduced into the center of the integrating sphere. The composites were measured alternately and at least 10 times in the same manner as the measurements of the UCNCs in the solution.

The UCNCs-PMMA composites are cylinders with the same diameter of 18 mm but different heights. The heights of the nanocrystals-PMMA composites for the un-doped references and for the samples containing the UCNCs vary between 9.8 nm and 11.8 mm. Unfortunately, the PMMA absorbs in the absorption range of the UCNCs. An absorption coefficient of $\alpha_{PMMA}$ = 0.015±0.001 mm$^{-1}$ at a wavelength of 1523 nm was calculated from reflectance and transmittance measurements of several pure PMMA cylinders. Additionally, the optical quality of the nanocrystals-PMMA composites varies and the color of the samples are anywhere from transparent to slightly yellowish. The variations in the optical quality were not considered in the calculations of the absorptance. Hence, the determination of the absorptance due to the UCNCs is the most uncertain parameter for calculating the internal UCQY of these samples. The absorptance of the UCNCs-PMMA composites was determined using the reference nanocrystals-PMMA composite with the closest concentration of un-doped nanocrystals to the concentration of the corresponding upconverter sample that simultaneously showed the most similar color and height to the UCNCs-PMMA composites.

# 5. Absorptance

The absorptance of the UCNCs-PMMA composites was calculated from transmittance measurements with a commercial spectrophotometer (Cary500i, Varian Inc.). The transmittance of the core-"giant" shell samples along with the reference sample, which were used for the calculations of the absorptance by the UCNCs, is shown in Figure S4a. The absorptance of the upconverter samples was calculated with respect to the corresponding un-doped reference sample using

$$A_{\text{sample}}(\lambda) = T_{\text{NC,ref}}(\lambda) - T_{\text{NC,sample}}(\lambda), \qquad (S8)$$

with the transmittance of the un-doped reference sample $T_{\text{NC,ref}}$ and the transmittance of the upconverter sample $T_{\text{NC,sample}}$. Different levels of parasitic absorption were observed for the different composites owing to the different optical quality and optical path length of the excitation beam through the composites. Consequently, the absorptance of the sample $A_{\text{sample}}$ includes parasitic absorptions, which should not be used to calculate the internal UCQY of the UCNCs. At a wavelength of 1450 nm no absorptance due to the $\beta$-NaYF$_4$: Er$^{3+}$ upconverter is expected. Therefore, the absorptance of the sample $A_{\text{sample}}$ was adjusted for an absorptance of zero at a wavelength of 1450 nm

$$A_{\text{UC}}(\lambda) = A_{\text{NC,smaple}}(\lambda) - A_{\text{NC,sample}}(1450 \text{ nm}) \qquad (S9)$$

The absorptance of the UCNCs in the composites $A_{\text{UC}}$ is shown in Figure S4b. The spectral resolution and the signal-to-noise ratio are both very low. However, the peak of $A_{\text{UC}}$ emerges at a wavelength of 1523 nm as expected and also observed in the excitation profiles shown in Figure 2a. Considering the accuracy of the spectrophotometer, which is commonly given with 1% absolute, the peak values of the absorptance by the upconverter from Figure S4b agree very well with values determined with the photoluminescence setup. All values are listed in Table S3.

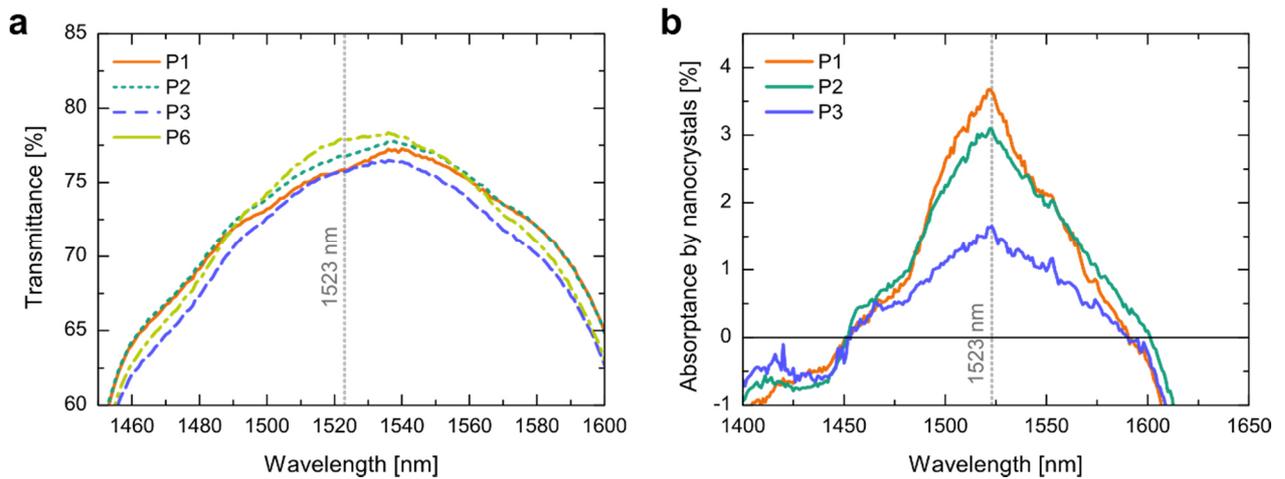

**Figure S4.** (**a**) Transmittance of the core-"giant" shell UCNCs-PMMA composite samples (P1,P2,P3) along with the transmittance of a reference with un-doped nanocrystals (P6). See Table S3 for sample abbreviations. (**b**) Although the spectral resolution is fairly low, the absorptance by the nanocrystals shows the expected spectral shape for $\beta$-NaYF$_4$: Er$^{3+}$ with a characteristic peak at a wavelength of 1523 nm.

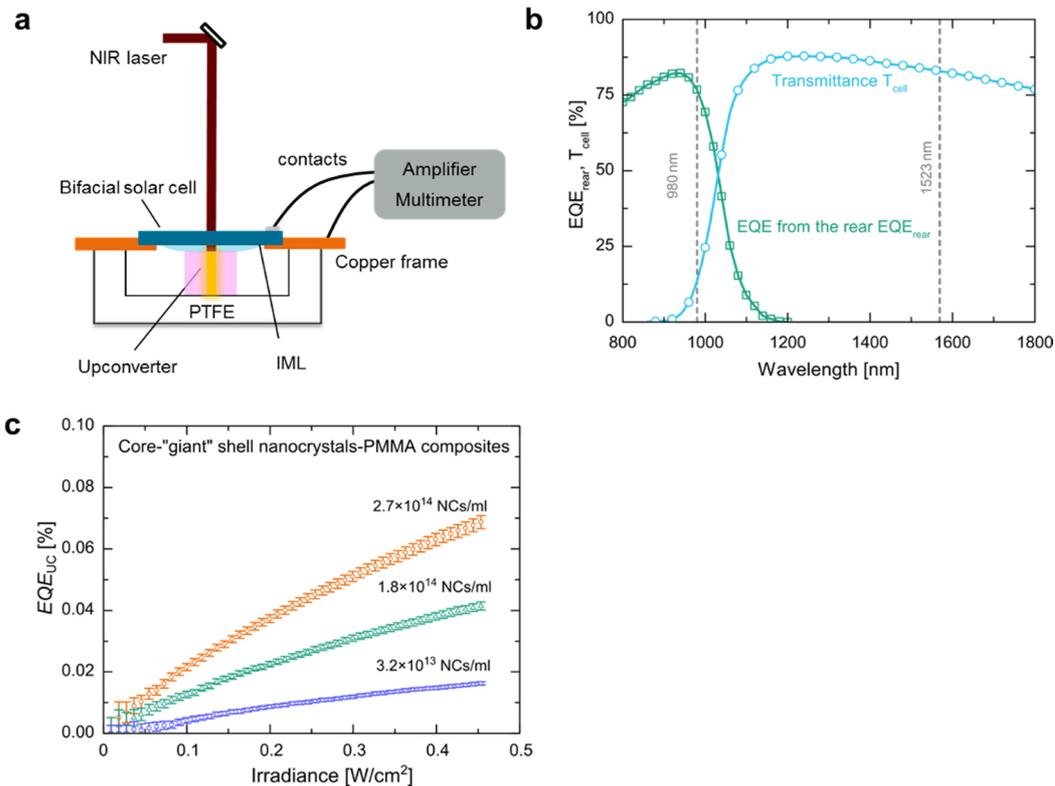

**Figure S5.** (**a**) Sketch of the setup to measure upconverter solar cell devices under monochromatic laser excitation. (**b**) The bifacial silicon solar cells have been optimized to feature a high transmittance $T_{cell}$ of sub-bandgap photons through the cell and a large EQE of the upconverted photons from the rear side $EQE_{rear}$. (**c**) External quantum efficiency of a bifacial silicon solar cell due to upconversion of sub-bandgap photons $EQE_{UC}$. Lower $EQE_{UC}$ were found for lower concentrations of core-"giant" shell UCNCs in PMMA.

# 6. Quenching Analyses (by Water and CHCl$_3$ vs CDCl$_3$)

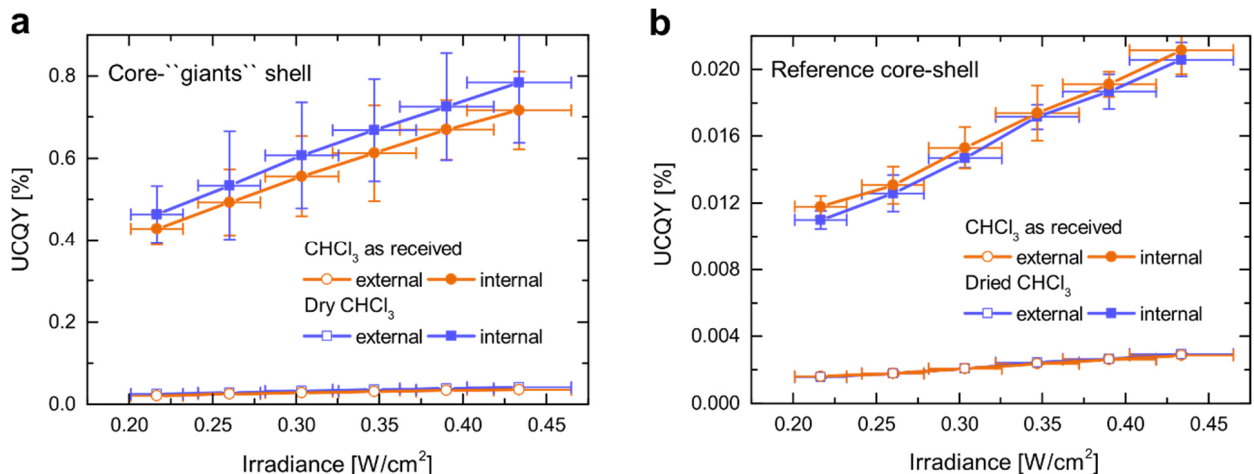

**Figure S5.** Quenching of the UCQY by water molecules was investigated for (**a**) the core-"giant" shell UCNCs and (**b**) another reference core-shell sample. This sample consisted of β-NaYF$_4$ UCNCs doped with 30% Er$^{3+}$ with a core diameter of 14.3 nm and a 3.2 nm thick β-NaLuF$_4$ shell. We compared the UCQY of the core-"giant" shell UCNCs and more commonly used reference core-shell UCNCs[1] in the solvent CHCl$_3$ as received as well as in dried CHCl$_3$. The dried CHCl$_3$ was obtained by adding 80 g of the as-received CHCl$_3$ into a flask containing 11.8 g of the molar sieve 4Å and slightly mixing the solution for 72 h to extract all water molecules from the CHCl$_3$. The same amount of UCNCs were precipitated and subsequently dispersed with the same amount of either the as received CHCl$_3$ or

---

[1] These reference core-shell UCNCs consist of a β-NaYF$_4$ core with doping of 30% Er$^{3+}$, core diameter 14.3±0.8 nm, and a β-NaLuF$_4$ shell, core-shell diameter 20.7±1.6 nm

the dried CHCl$_3$. For each irradiance value the UCQY of every sample was measured 10 times. Both samples show no significant difference in the external and internal UCQY when they are dispersed in dried CHCl$_3$ or in the as received CHCl$_3$.

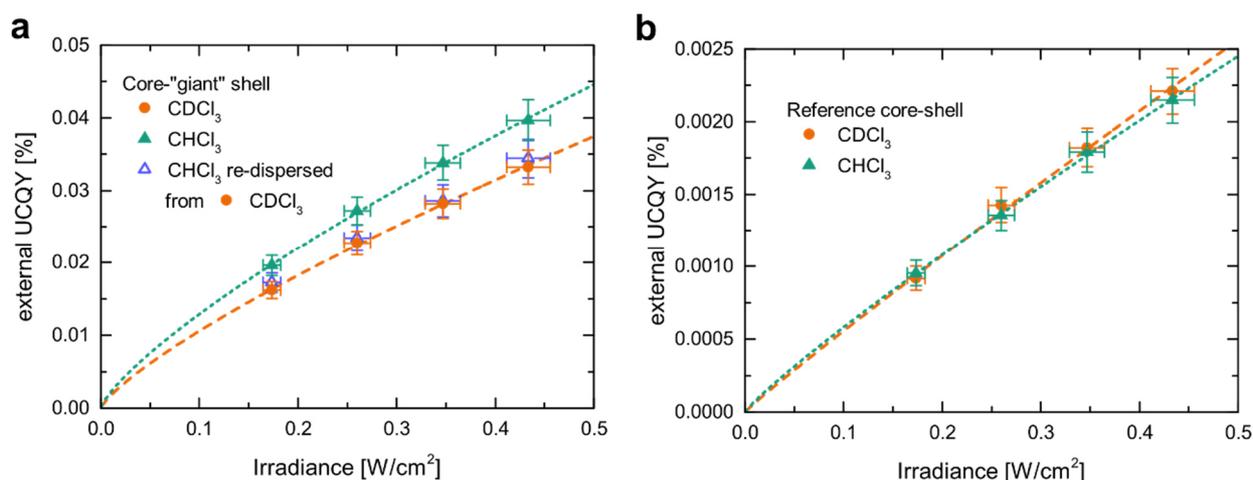

**Figure S6.** Quenching due to the C-H vibrational oscillation in the solvent CHCl$_3$ was investigated for (**a**) the core-"giant" shell UCNCs and for (**b**) the reference core-shell UCNCs. As previously discussed for water molecules as quenchers, the same amount of core-"giant" shell UCNCs and the reference core-shell UCNCs were dispersed in CHCl$_3$ and CDCl$_3$. Again, the external UCQY was measured for every sample and for each irradiance value 10 times. No significant quenching due to the C-H vibrational oscillation of the CHCl$_3$ was found. We attribute the lower external UCQY of the core-"giant" shell UCNCs in CDCl$_3$ to a slightly lower concentration of the nanocrystals in the solution. In an additional measurement (blue open triangle), where the CDCl$_3$ sample was re-dispersed in CHCl$_3$, nearly the same external UCQY was determined as before in CDCl$_3$.

## 7. Energy Migration (EM) of the Donor Excitation

For the different literature values, as listed in Table 1, the probability of migration $P_{migrate}$ as a function of the number of hopping processes was calculated for different irradiances. The results for the data reported by Tkachuk et al. are shown in Figure S7a. Tkachuk et al. investigated the material system LiYF$_4$: Er$^{3+}$, which is very similar to NaYF$_4$ Er$^{3+}$ that was used in this work.[4] The probability $P_{migrate}$ as a function of the average migration distance $R_{migrate}$ is shown Figure S10b. The values were calculated from the data shown in Figure S7a using Equations 11, 12 and 18. The probability distribution of the $R_{migrate}$ values can be determined by the derivation of $P_{migrate}$ as a function $R_{migrate}$. The results for the data from Tkachuk et al.[4] and Agazzi et al.[5] are shown in Figure S10c and Figure S10d. The weighted average values of $R_{migrate}$ are all in the range of several nanometres also depending on the irradiance, as given in Table S5 for irradiances of 0.5 W/cm$^2$ and 100 W/cm$^2$. The values from first simple approximation using the hopping time $\tau_0$ are much larger than the ones determined with the rate equation model.

The lowest value of $R_{migrate}$ was determined to be 3.1 nm using the data reported in ref. [4] and an irradiance of the excitation of 100 W/cm$^2$. Although this is our most conservative approximation, it means nearly 70% of the excitation energy generated in the active core with a radius of 9.6 nm, such as for the core-"giant" shell UCNCs, reaches the surface on its migration path. However, even if the donor excitation reaches the surface it does not mean that the excitation actually will be quenched because the surface quenching can be understood as another rate (probability) – with finite value - that adds into the rate equations given in Equation 11 The quenching will be dominant when the probability of the quenching process is larger than the other probabilities. The same applies for defects in the volume, which appear to be very crucial in this picture with rather large migration path length.

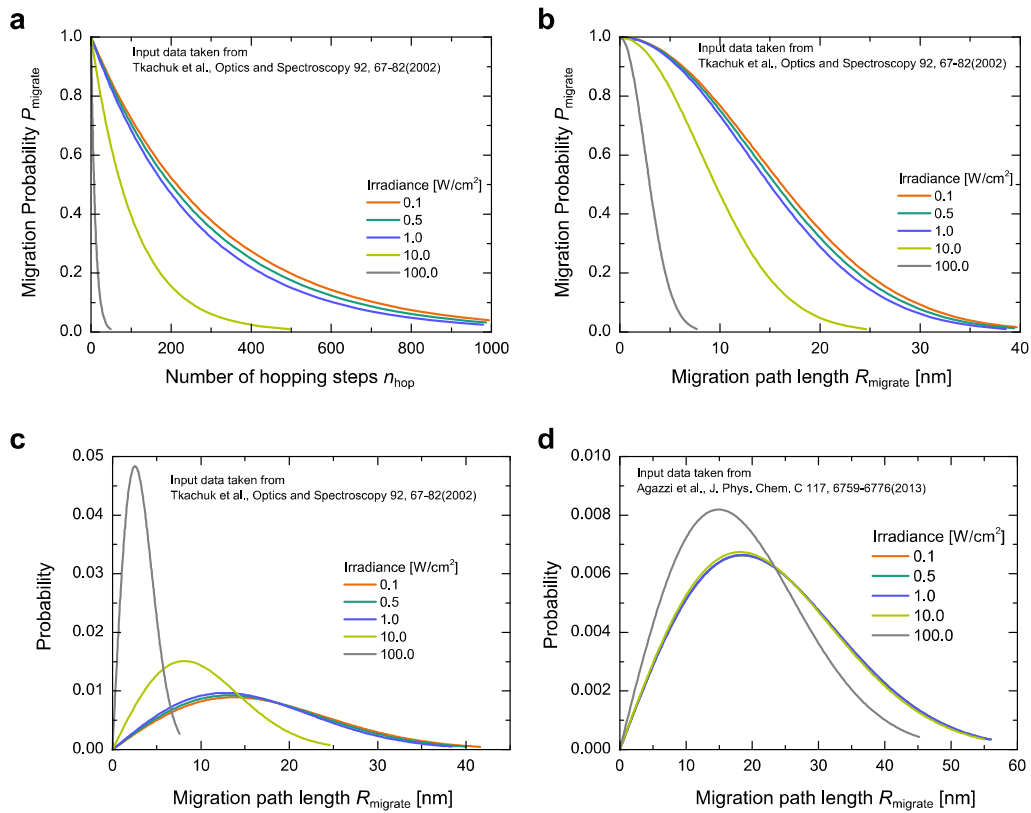

**Figure S7.** (**a**) Hopping processes from one rare-earth ion to another rare-earth ion are very efficient in the material systems under consideration. Using a simplified hopping rate equation model we determined a decreasing number of hopping steps $n_{hop}$ with increasing irradiance of the excitation. (**b**) Average migration path lengths $R_{migrate}$ were calculated from the number of hopping steps using a 3D random walk model. $P_{migrate}$ describes the probability an excitation energy can be found after $n_{hop}$ or $R_{migrate}$ (**c,d**) The probability distributions of the $R_{migrate}$ were calculated for different data sets, which are reported in the literature. The weighted average values of these probability distributions are given in Table S5.

**Table S4.** Number of hopping steps $n_{hop}$ and the corresponding average migration path length $R_{migrate}$, which was estimated using the average time an $Er^{3+}$ excitation resides before it is hopping to another $Er^{3+}$ $\tau_0$ as well as a more sophisticated analysis using a rate equation hopping model. A strong dependence of $n_{hop}$ and consequently of $R_{migrate}$ on the irradiance of the excitation $I$ was determined with the rate equation hopping model. The values as calculated from the data reported by Tkachuk *et al.* were further considered because these are the most conservative ones and the material system $LiYF_4$: $Er^{3+}$ is very much alike $NaYF_4$: $Er^{3+}$, which was used in this work.

| $\tau_0$ [µs] | $n_{hop}$ | $R_{migrate}$ [nm] | $n_{hop}$ for $I$=0.5 Wcm$^{-2}$ ($I$=100 W/cm$^2$) | $R_{migrate}$ [nm] for $I$=0.5 Wcm$^{-2}$ ($I$=100 W/cm$^2$) | Calculated from Reference |
|---|---|---|---|---|---|
| 1.8 | 5580 | 83 | - | - | 6 |
| 3.1 | 2890 | 59 | 249(10) | 15.7 (3.1) | 4 |
| 1.5 | 5140 | 79 | 526(346) | 22.8 (18.4) | 5 |

**Further considerations of EM and quenching.** From the analysis of the static quenching and the EM one can understand why surface related quenching plays an important role in the achievable UCQY of UCNCs. Furthermore, we can also understand why the increase in the nanocrystal size as well as a sufficiently thick passivation shells increase the UCQY considerably. From the long migration path length it becomes clear that lattice defects in the active core are very destructive and prevention or healing of these defects is mandatory to obtain high UCQY. Bulk materials are typically annealed at high temperatures far above 500 °C for several hours, which may help to remove lattice defects in the volume, whereas nanomaterials are typically synthesized at only 300 °C for one hour. The huge difference in temperature and time between the nanomaterials and the bulk materials may also lead to different distribution of the rare-earth ions at their lattice positions. During the synthesis, the core-"giant" shell UCNCs have

## 8. Comparison with Values found in Literature

UCQY values for nanomaterials and bulk materials are summarized in Table S5. In the literature, the irradiances used to determine the UCQY are typically 2 order of magnitude higher than the ones used in this work and also above state-of-the-art solar concentrator systems, as discussed in the paper. All values in Table S6 were taken from the references. Only the values given by Shao et al. in ref. [7] had to be estimated because they give a power conversion efficiency and not an UCQY.

Shao *et al.* reported an energy conversion efficiency of 3.9% using an irradiance of 18 W/cm$^2$ at 1523 nm for $\beta$-NaYF$_4$: 10% Er$^{3+}$ / NaYF$_4$ core/shell nanoparticles.[7] This energy efficiency can be converted in a UCQY by considering the energy of the photons ($E_\lambda = hc/\lambda$) in the emission and excitation spectra, where $h$ is Planck's constant, $c$ is the speed of light, and $\lambda$ the wavelength. For the data given in the paper, we estimated an UCQY of 1.7% at 1532 nm and 18W/cm$^2$ using the equation

$$UCQY = \frac{E_{\lambda,excitation}}{E_{\lambda,emission}} 3.9\% \approx 1.7\% . \qquad (S26)$$

The corresponding normalized internal UCQY is 1.7%/(18W/cm$^2$) = (0.017/18) cm$^2$/W = 0.00094 cm$^2$/W. Furthermore, in ref. [7] the UC energy efficiency was determined with power meters and additional optical filter. The IR photons from the excitation were measured using an 1100 nm long-pass filter. The problem with such a method is that the emission of the $^4I_{13/2}$ to $^4I_{15/2}$ around 1550 nm is also detected when the excitation signal is measured. As a result, the sample's absorptance can be highly underestimated and therefore the UCQY overestimated, as reported by MacDougall et al. [8]. MacDougall et al. reported an overestimation factor of ~2.5× for 10% Er$^{3+}$ doping in NaYF$_4$ due to detected emission from $^4I_{13/2}$ to $^4I_{15/2}$, which means the UCQY value in ref. [7] would be around 0.7%, as listed in Table S6.

Some references in Table 6S are marked with (R), which means that the UCQY values in these references were determined with a relative measurement technique by using a reference sample with precisely known QY. Such a method is known to be error-prone in the near-infrared because reliability issues of the absolute QY values for the typically used reference dyes.[9]

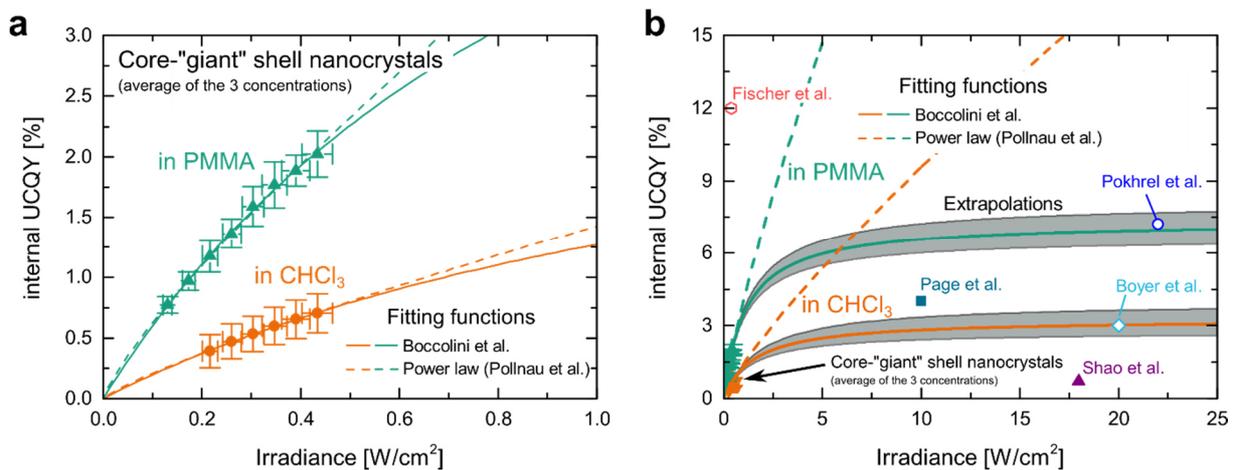

**Figure S8.** Extrapolations of the internal UCQY as a function of the irradiance. We used two different models, a more conservative one adapted from Boccolini et al. [10] and an optimistic one using the power law for the upconversion luminescence from Pollnau et al.[11]. The power law does not account for saturation and consequently smaller slopes at higher irradiances, which results in an overestimation of the UCQY for extrapolated values. (**a**) In the low irradiance regime, both model can be used to describe the experimentally determined internal UCQY of the core-"giant" shell UCNCs in solution of CHCl$_3$ and embedded in PMMA very well. (**b**) For higher irradiances, it is very difficult to predict UCQY values and both models differ significantly. We expect the reality to be the somewhere between the two models. However, even for the conservative model higher UCQY values can be estimated for the core-"giant" shell UCNCs than for bulk materials. For example, the internal UCQY as reported by Page et al. [12], Boyer and van Veggel [3], Pokhrel et al. [13], and Fischer et al.[14] are shown. Furthermore, the best comparable UCNCs reported by Shao et al. [7] are depicted as well. All UCQY values can also be found in Table S6.

**Table S5. Overview over UCQY values reported in the literature.** The used excitation wavelength $\lambda_{exc}$, the dominant emission wavelength $\lambda_{em,main}$, the irradiance $I$, the internal UCQY, and the corresponding figure of merit – the normalized internal UCQY – are given.

| Material core / shell | Size core (+shell) [nm] | $\lambda_{exc}$ [nm] | $\lambda_{em,main}$ [nm] | Irradiance $I$ [W/cm$^2$] | Internal UCQY [%] | Normalized internal UCQY [cm$^2$/W] | Ref. |
|---|---|---|---|---|---|---|---|
| $\beta$-NaYF$_4$: 28% Er$^{3+}$ / $\beta$-NaLuF$_4$ | 19.2(38.8) | 1523 | 980 | 0.43±0.03 | 0.71±0.08 | 0.017±0.002 | This work |
| **$\beta$-NaYF$_4$: 28% Er$^{3+}$ / $\beta$-NaLuF$_4$ in PMMA** | **19.2(38.8)** | **1523** | **980** | **0.43±0.03** | **2.01±0.19** | **0.047±0.005** | **This work** |
| $\beta$-NaYF$_4$: 10% Er$^{3+}$ / $\beta$-NaYF$_4$ | 22 (25/38) | 1523 | 655# | 18 | 0.7* | 0.00039* | 7 |
| LiYF$_4$: 10% Er$^{3+}$ | 85 | 1490 | 975 | 150 | 1.2±0.1 | 0.00008* | 15 (R) |
| $\beta$-NaYF$_4$: 20% Yb$^{3+}$, 2% Er$^{3+}$ | 100 | 980 | 540 | 150 | 0.30±0.1 | 0.00002* | 3 |
| $\beta$-NaGdF$_4$: 20 Yb$^{3+}$, 2% Er$^{3+}$ / $\beta$-NaYF$_4$ | 10.5(20.1) | 980 | 540 | 50 | 0.89 | 0.000178* | 16† |
| $\beta$-NaYF$_4$: 25% Yb$^{3+}$, 0.003% Tm$^{3+}$ / $\beta$-NaYF$_4$ | 30(42) | 980 | 800 | 78 | 3.5 | 0.00045* | 17 (R) |
| LiLuF$_4$: 20% Yb, 1% Er / LiLuF$_4$ | 28.0(50.7) | 980 | 540 | 127 | 5.0 | 0.0004* | 18 † |
| LiLuF$_4$: 20% Yb, 0.5% Tm / LiLuF$_4$ | 28.0(50.7) | 980 | 800 | 127 | 7.6 | 0.0006* | 18 † |
| **$\beta$-NaYF$_4$: 25% Er$^{3+}$** | **bulk** | **1523** | **980** | **0.40±0.02** | **12.0±1.0** | **0.297±0.028** | **14** |
| $\beta$-NaYF$_4$: 20% Yb$^{3+}$, 2% Er$^{3+}$ | bulk | 980 | 822 | 22±3 | 7.2±1.2 | 0.004* | 13 |
| $\beta$-NaYF$_4$: 18% Yb$^{3+}$, 2% Er$^{3+}$ | bulk | 980 | 655 | 10 | 4.0 | 0.004* | 12 |
| $\beta$-NaYF$_4$: 20% Yb$^{3+}$, 2% Er$^{3+}$ | bulk | 980 | 540 | 20 | 3.0±0.3 | 0.002* | 3 |

*This value was estimated based on the data given in the corresponding reference.
†No or only very few details for the UCQY determination and method are given.
#Emission spectra have not been corrected for spectral sensitivity of the experimental setup
(R) Relative measurements to determine the UCQY, which is known to be error-prone in the near-infrared due to the questionable absolute QY of the used reference dyes.[9]

**Comparison of upconverter solar cell devices:** To compare the upconverter solar cell device performance of the bulk material with the UCNCs-PMMA composites, the absorptance of the bulk needs to be related to the one of the UCNC sample. In this work, an absorptance of only 3.8% was reached for the UCNCs-PMMA sample with the highest UCNCs concentration. Much higher absorptance values can be achieved, by increasing the density of nanocrystals in the PMMA. Scaling the absorptance of the UCNCs-PMMA composite to the bulk value of ~66% [14,19], we can estimate an $EQE_{UC}$ = 66% / 3.8% × 0.068% = 1.18 %. The corresponding bulk $EQE_{UC}$ value is around 5.7%. This means that the UCNCs in a photovoltaic device are only 1/5 less efficient than the values achieved with the very best bulk materials available. This is again in excellent agreement with the optical UCQY measureemtns presented in this work.

We can also compare our upconverter solar cell device results to the state of the art devices in the literature. Upconverter nanocrystals have been applied to dye sensitized solar cells [20] and organic solar cells [21]. However, as said before we are first to report a quantitative measurement for crystalline silicon solar cells. When we compare our results with those reported in the literature, our device is a factor 116x better than the one using a dye-sensitized solar cell [20] and 169x better than the one using an organic solar cell [21]. Here, we detail the typically used figure of merit (FOM) which is the normalized $EQE_{UC}$.

**Table S6.** Overview over highest $EQE_{UC}$ values reported in the literature for UCNCs photovoltaic devices for different solar cell technologies.

| Solar cell technology | Irradiance [W/cm$^2$] | $EQE_{UC}$ [%] | Normalized $EQE_{UC}$ (FOM) [10$^{-4}$ cm$^2$/W] | References |
|---|---|---|---|---|
| c-Si | 0.45 | 0.068 | 15.1 | This work |
| Dye-sensitized | 8 | 0.011 | 0.13 | [20] |
| Organic | 4.9 | 0.004 | 0.089 | [21] |